\documentclass[journal,twoside]{IEEEtran}
\usepackage{cite}               
\usepackage{graphicx}           
\usepackage{amsmath,amssymb,bm} 
\usepackage{nccmath}            
\usepackage{float}
\usepackage{tikz,circuitikz}    
\usepackage{subfig}             
\usetikzlibrary{calc,decorations.markings}
\usepackage{xcolor}
\usepackage{siunitx}            
\usepackage{url}
\usepackage{hyperref}
\usepackage{stfloats}
\usepackage{cuted}
\usepackage{chngcntr} 
\usepackage{enumitem}

\newcommand{\eqsref}[1]{Eq.~\eqref{#1}} 
\title{Exact Local-Field Renormalization for Deep-Subwavelength Particles in Rectangular Cavities}

\author{Koffi‑Emmanuel~Sadzi,~\IEEEmembership{Student Member,~IEEE,}
        and~Yakir~Hadad,~\IEEEmembership{Senior Member,~IEEE}
\thanks{Manuscript received XXX; revised XXX. (Corresponding author: Y. Hadad.)}%
\thanks{The authors are with the School of Electrical Engineering, Tel‑Aviv University, Tel‑Aviv 69978, Israel (e‑mail: yakirhad@tauex.tau.ac.il).}%
\thanks{Color versions of one or more figures in this article are available online at https://ieeexplore.ieee.org.}%
\thanks{Digital Object Identifier XX.XXXX/TAP.2025.XXXXXXX}}%

\begin{document}
\maketitle

\begin{abstract}
We present a rigorous, semi-analytical framework for predicting the joint eigenfrequencies of a deep-subwavelength particle embedded in a perfectly conducting rectangular cavity. The formulation retains the \emph{full} cavity-mode spectrum and is therefore strictly causal, in contrast to Jaynes–Cummings–type models that truncate the spectrum and fail in strong coupling. A ladder-type Green-function renormalization is introduced: three successive subtractions—cavity minus rectangular waveguide, waveguide minus parallel plate, and parallel plate minus free space—remove the “$\infty-\infty$’’ singularity of the local field. The resulting local dyadic Green function is obtained with a rapidly convergent recursive algorithm whose cost scales linearly with the number of spectral terms. Once the local field is known, the cavity-renormalized polarizability
\(
\boldsymbol{\alpha}_{\mathrm{eff}}(\omega)=\bigl[\boldsymbol{\alpha}^{-1}-\mathbf
G_{\mathrm{loc}}(\mathbf r')\bigr]^{-1}
\)
yields the coupled resonances from
\(
\det\!\bigl[\boldsymbol{\alpha}_{\mathrm{eff}}^{-1}(\omega)\bigr]=0.
\)
Benchmark cases involving isotropic, gyrotropic, and chiral spheres confirm exponential convergence and capture both weak- and strong-coupling regimes without adjustable parameters. The method is numerically robust, applies to arbitrary material tensors, and can be extended to structured waveguides whose transverse eigenfunctions are obtained numerically, providing a practical design tool for cavity–particle systems from microwave to terahertz frequencies.
\end{abstract}

\begin{IEEEkeywords}
Cavity QED, dyadic Green’s function, strong and ultra‑strong coupling, ladder‑type subtraction, plasmonic nanoparticles, chiral media.
\end{IEEEkeywords}
\IEEEpeerreviewmaketitle


\section{Introduction}\label{sec:introduction}

Light--matter coupling underpins numerous modern photonic technologies.  
One of the most effective methods to control this interaction involves engineering the local photonic environment—often by embedding emitters within metallic or dielectric cavities~\cite{Plasmon_Induced_Circular_Govorov,Giant_circular_dichroism_Govorov} or optical microcavities~\cite{LiuLiXiao2017_EIT,mann2018manipulating,Armani_single_molecule,huang2011electromagnetically,mann2022topological}.  
A prominent manifestation of this approach is the Purcell effect, where spontaneous‐emission rates are enhanced or suppressed by modifying the local density of states through resonant~\cite{purcell1946,kleppner1981inhibited,hulet1985inhibited,yablonovitch1987inhibited,gabrielse1985observation,englund2005controlling,hadad2020possibility,chiriaco2023thermal} or reflective~\cite{Sipe1984,Sipe1985} boundaries.  
These cavity‐quantum‐electrodynamic (CQED) principles are foundational to advances in molecular‐reaction control~\cite{Armani_single_molecule,Mukamel_cavity_spectroscopy,George2016_Multiple_Rabi_Splittings,Schwartz_rabi_split,Flick2018,Ribeiro2018,KenaCohen2019}, quantum sensing, and quantum information processing~\cite{Degen_Quantum_sensing,Zhang_Strongly_Coupled_Magnons,Reiserer_Cavity_based_quantum_networks,Ladd2010_Quantum_computers}.

When the emitter’s intrinsic lifetime \(\tau_{p}\) exceeds the cavity round‐trip time \(\tau_{B}=d/c\), the emitted radiation interacts repeatedly with the cavity walls, leading to the strong‐coupling regime.  
Traditionally, this regime is described using the Jaynes–Cummings model and its rotating‐wave approximation~\cite{Gerry_Knight_Book,JC_book}, suitable for weak or moderate coupling~\cite{George2016_Multiple_Rabi_Splittings,Schwartz_rabi_split}.  
At stronger coupling levels, however, such simplified models fail and can even violate causality~\cite{JC_book,NatComm_Causality}.

These issues can be rigorously addressed by employing the medium Green’s function (GF), which inherently captures the full spectral response of the system.  
However, this exact approach introduces mathematical complexity because the GF’s singular term must be precisely subtracted to evaluate the \emph{local‐field} GF,
\[
\mathbf{G}_{\text{loc}}(\mathbf{r}) = \mathbf{G}(\mathbf{r},\mathbf{r}) - \mathbf{G}_{\text{FS}}(\mathbf{r},\mathbf{r}).
\]
While such extraction is straightforward near open or reflective boundaries~\cite{Sipe1984,Sipe1985}, it becomes significantly challenging in closed cavities.  
In cavity QED, this local‐field GF (also termed the \emph{renormalized} or self‐extracted dyadic GF) governs critical observables: its imaginary part sets the spontaneous‐decay rate, and its real part determines the Lamb shift.  
Conventional evaluations rely on approximations such as mode truncation or point splitting, both of which lack rigor and numerical stability.

In this paper, we propose an \emph{exact and numerically stable} renormalization formalism, together with an efficient computational methodology, to evaluate mutual coupling between a resonant particle and a cavity for arbitrary coupling strengths.  
The particle response is characterized by its polarizability tensor \(\boldsymbol{\alpha}(\omega)\), encapsulating both material and geometric properties.  
Within classical and semi‐classical frameworks, this tensor relates the induced dipole moment to the \textit{local field}—the field at the particle’s location excluding its self‐contribution.  
In cavity configurations, the local field necessarily includes significant back‐scattered waves from the cavity boundaries.

To compute the local field rigorously, the dyadic GF in bounded media must be decomposed into clearly defined \textit{primary} (free‐space) and \textit{secondary} (cavity‐induced) components.  
Although conceptually straightforward, such decomposition often poses severe numerical difficulties.  
Even though the GF’s singularity near a dipole source is formally identical in bounded and unbounded scenarios, the two mathematical forms differ enough to make direct subtraction unstable.

Here, we introduce a robust numerical strategy—a \textit{ladder‐type recursive subtraction method}—for accurate derivation of the local GF in rectangular cavities.  
Our approach leverages alternative GF representations~\cite{felsen2001radiation}, which are usually used for field computation, to simplify the decomposition in geometries amenable to the separation of variables and lead to efficient local field GF regularization.

The formalism is general and broadly applicable, readily extending to electric, magnetic, and chiral (bi‐anisotropic) particles.  
Additionally, the technique can be adapted for waveguide regularization and extended to geometries separable in cylindrical or spherical coordinates.  
Using our method, we examine several particle‐cavity systems exhibiting strong coupling across various cavity modes and particle types, including isotropic, anisotropic gyrotropic, and bi‐isotropic chiral resonators.

The manuscript is organised as follows.  
Section~\ref{sec:math} establishes the joint cavity–particle eigenvalue problem, introducing the polarizability formalism, the local Green dyadic, and the recursive ladder-subtraction scheme.  
Section~\ref{sec:localGF} applies this scheme—using the alternative-Green-function method of Felsen and Marcuvitz \cite{felsen2001radiation}—to derive closed-form expressions for the local dyadic Green function excited by a \(z\)-polarised electric point current (the magnetic analogue is obtained in Appendix~\ref{sec:level_3b}).  
Section~\ref{sec:stability_efficiency} verifies numerical stability and demonstrates the expected exponential convergence.  
Section~\ref{sec:level_gen} then generalizes the recipe, showing how to construct the complete dyadic Green function for an arbitrarily oriented electric or magnetic source.  
Finally, Section~\ref{sec:level_ex} illustrates the technique with eigenfrequency calculations for isotropic, gyrotropic, and chiral particles embedded in the cavity.

\section{Mathematical Methodology}\label{sec:math}

\subsection{Discrete Dipole Approximation}

\begin{figure}[tb]
  \centering
  \includegraphics[width=0.48\textwidth]{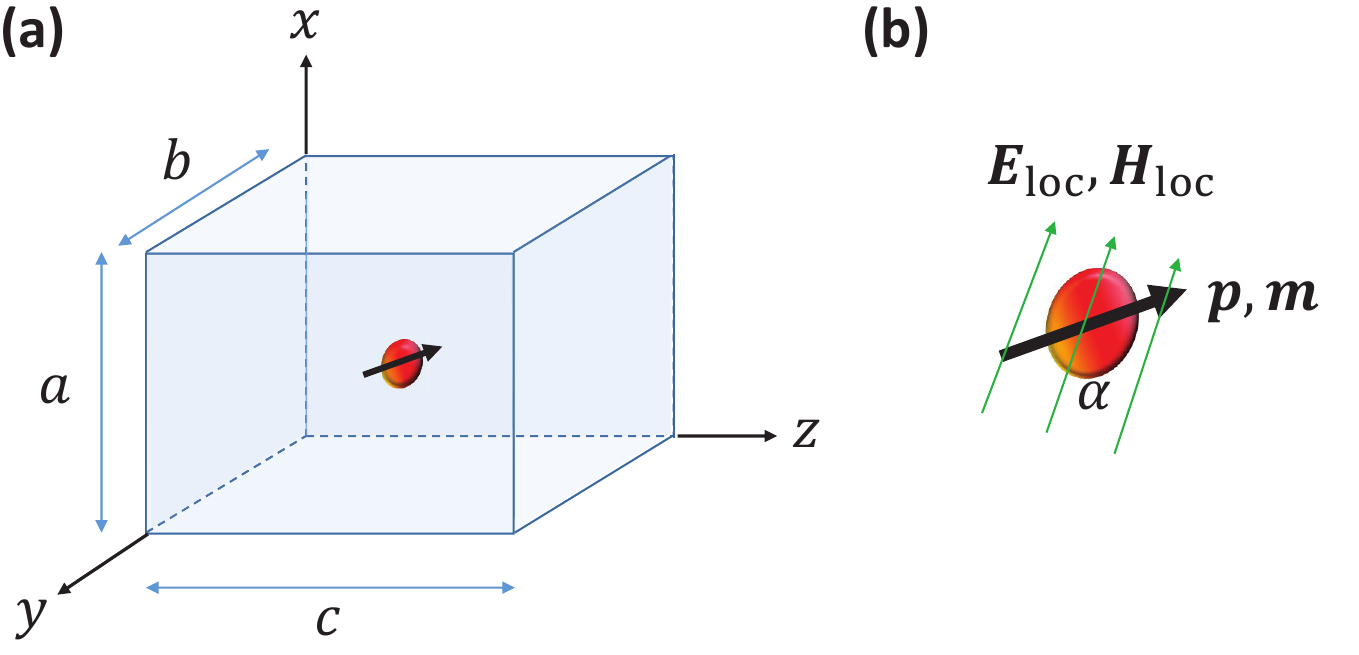}
  \caption{%
    Problem geometry.
    (a)  A resonant (possibly bi‐anisotropic) particle at \(\mathbf{r}'=(x',y',z')\) is embedded in a rectangular PEC cavity of size \(a\times b\times c\).
    (b)  The induced dipoles \(\mathbf{s}=[\mathbf{p};\mathbf{m}]\) are driven by the local field \(\mathbf{U}_{\mathrm{loc}}=[\mathbf{E}_{\mathrm{loc}};\mathbf{H}_{\mathrm{loc}}]\) through the polarizability tensor \(\boldsymbol{\alpha}\).
  }
  \label{fig:setup}
\end{figure}

The geometry under study is sketched in Fig.~\ref{fig:setup}.  
A deep–subwavelength resonant particle is placed inside a perfectly conducting rectangular cavity.  
Our goal is to find the coupled eigenfrequencies for arbitrary particle–cavity coupling, including bi–anisotropic particles.

Because the particle size is much smaller than the cavity dimensions, a volume‐integral formulation is most appropriate \cite{PRR,novotny2012principles}.  
For a deep-subwavelength inclusion, higher–order multipoles can be neglected and the particle is fully described by an electric–magnetic dipole vector  
\(
\mathbf{s}
     =
     \bigl[
       \mathbf{p}\;;\;\mathbf{m}
     \bigr]^{\mathsf{T}},
\)
leading to the discrete–dipole approximation (DDA) \cite{Tretyakov2011,novotny2012principles}.  
The dipole moments respond to the \emph{local} field,
\begin{equation}
\mathbf{s}
  = \boldsymbol{\alpha}\,
    \mathbf{U}_{\mathrm{loc}},
\qquad
\mathbf{U}_{\mathrm{loc}}
  =
  \bigl[
    \mathbf{E}_{\mathrm{loc}}\;;\;
    \mathbf{H}_{\mathrm{loc}}
  \bigr]^{\mathsf{T}},
\label{eq:sU}
\end{equation}
where the $6\times6$ polarizability tensor  
\(
\boldsymbol{\alpha}
=
[
 \boldsymbol{\alpha}_{ee},  \boldsymbol{\alpha}_{em};
 \boldsymbol{\alpha}_{me},  \boldsymbol{\alpha}_{mm}
]
\)
captures both electric and magnetic, as well as cross-coupling, responses.

For a homogeneous host medium \((\varepsilon_{0},\mu_{0})\) the static polarizability \(\boldsymbol{\alpha}_{s}\) must be augmented by the universal radiation-correction term that enforces energy conservation \cite{Tretyakov2011}:
\begin{equation}
\boldsymbol{\alpha}^{-1}
  = \boldsymbol{\alpha}_{s}^{-1}
  + j\frac{k^{3}}{6\pi}
    \begin{bmatrix}
      \tfrac{1}{\varepsilon_{0}}\mathbf{I}_{3} & \mathbf{0}\\
      \mathbf{0} & \tfrac{1}{\mu_{0}}\mathbf{I}_{3}
    \end{bmatrix},
\label{eq:radcorr}
\end{equation}
with \(k=\omega/c\) and \(\mathbf{I}_{3}\) the $3\times3$ identity dyad. Second–order (\(\propto k^{2}\)) corrections to the \emph{real} part of the polarizability may also be included, but—unlike the \(k^{3}\) term in the imaginary part—they introduce no new physics, merely producing a slight, volume–dependent resonance shift \cite{bohren_huffman}.

%

\subsection{Local Field, Effective Polarizability, and Eigenproblem}
\label{sec:effpol}

\subsubsection{Local field}

Any Green’s dyad in a structured medium can be split into a free–space singular part and a regular scattering part.  
For the cavity (CV) we write
\begin{equation}
\mathbf{G}_{\!\mathrm{CV}}(\mathbf{r},\mathbf{r}')
  = \mathbf{G}_{\!\mathrm{FS}}(\mathbf{r},\mathbf{r}')
  + \mathbf{G}^{s}(\mathbf{r},\mathbf{r}'),
\label{eq:GFdecomp}
\end{equation}
where all dyads are $6\times6$ and the superscript \(s\) denotes the secondary (source-free) contribution.  
Defining fields through  
\(\mathbf{U}=j\omega\mathbf{G}\,\mathbf{s}\),  
the local field at the particle position \(\mathbf{r}'\) is  
\begin{equation}
\mathbf{U}_{\mathrm{loc}}(\mathbf{r}')
  = j\omega\mathbf{G}_{\mathrm{loc}}(\mathbf{r}')\,\mathbf{s}
  + \mathbf{U}_{\mathrm{in}}(\mathbf{r}'),
\label{eq:Uloc}
\end{equation}
with the \emph{local} Green’s dyad
\begin{equation}
\mathbf{G}_{\mathrm{loc}}(\mathbf{r}')
 = \mathbf{G}^{s}(\mathbf{r}',\mathbf{r}').
 \label{eq:sec to local}
\end{equation}
This regularized dyad dictates both the Lamb shift (its real part) and the decay rate (its imaginary part) of the resonator.

\subsubsection{Effective polarizability}

Eliminating \(\mathbf{U}_{\mathrm{loc}}\) between \eqsref{eq:sU} and \eqsref{eq:Uloc} yields
\begin{equation}
\mathbf{s}
 = \underbrace{\Bigl[
     \boldsymbol{\alpha}^{-1}
     - j\omega\mathbf{G}_{\mathrm{loc}}(\mathbf{r}')
   \Bigr]^{-1}}_{\displaystyle\boldsymbol{\alpha}_{\mathrm{eff}}}
   \mathbf{U}_{\mathrm{in}}(\mathbf{r}'),
\label{eq:alphaeff}
\end{equation}
defining the \emph{effective} polarizability \(\boldsymbol{\alpha}_{\mathrm{eff}}\).  
In free space \(\mathbf{G}_{\mathrm{loc}}=0\), so \(\boldsymbol{\alpha}_{\mathrm{eff}}=\boldsymbol{\alpha}\); inside the cavity both detuning and broadening are altered by \(\mathbf{G}_{\mathrm{loc}}\).

\subsubsection{Eigenvalue condition}

For the self‐sustained (source-free) modes of the coupled cavity–particle system we set \(\mathbf{U}_{\mathrm{in}}=0\).  
Non-trivial solutions of \eqsref{eq:alphaeff} therefore exist only when
\begin{equation}
\det\!\Bigl[\boldsymbol{\alpha}_{\mathrm{eff}}^{-1}(\omega)\Bigr]=0,
\label{eq:eig}
\end{equation}
which gives the sought eigenfrequencies.

\subsection{Numerical Evaluation of \(\mathbf{G}_{\mathrm{loc}}\):
           Ladder‐Subtraction Scheme}

Directly computing  
\(
\mathbf{G}_{\mathrm{loc}}
 =\mathbf{G}_{\!\mathrm{CV}}
 -\mathbf{G}_{\!\mathrm{FS}}
\)
at \(\mathbf{r}=\mathbf{r}'\) is numerically unstable:  
\(\mathbf{G}_{\!\mathrm{CV}}\) is a double series over discrete cavity modes, whereas \(\mathbf{G}_{\!\mathrm{FS}}\) is a double integral over a continuum, leading to an “\(\infty-\infty\)” subtraction.

\subsubsection{Existing remedies}

Image theory or Ewald-type summations can in principle regularize the difference, but both approaches become cumbersome for off-center particles, high‐order modes, or large cavities, and convergence is typically slow \cite{Sheng1986,Capolino2007,Capolino2008,Nam1998,PRA2023}.  
Perturbative methods also fail once strong coupling or multimode interactions are involved.

\subsubsection{Ladder-subtraction strategy}

We instead introduce two intermediate Green’s dyads—those of an infinite rectangular waveguide (RG) and an infinite parallel‐plate waveguide (PP)—and split the secondary dyad into three \emph{well‐behaved} differences:
\begin{align}
\mathbf{G}^{s}(\mathbf{r},\mathbf{r}')
&= \underbrace{\bigl[\mathbf{G}_{\!\mathrm{CV}}
                     -\mathbf{G}_{\!\mathrm{RG}}\bigr]}_{\text{Sub.\,(I)}} \nonumber\\
&\quad+\underbrace{\bigl[\mathbf{G}_{\!\mathrm{RG}}
                         -\mathbf{G}_{\!\mathrm{PP}}\bigr]}_{\text{Sub.\,(II)}}
 +\underbrace{\bigl[\mathbf{G}_{\!\mathrm{PP}}
                    -\mathbf{G}_{\!\mathrm{FS}}\bigr]}_{\text{Sub.\,(III)}}.
\label{eq:ladder}
\end{align}
Using the theory of alternative GF representation \cite{felsen2001radiation}, for each bracketed term we choose \emph{matching modal bases}, so that the subtraction occurs inside the common mode expansion. 
Consequently, every difference involves only regular source-extracted one-dimensional Green’s functions and converges exponentially fast with the spectral integration/summation parameter.
The detailed application of \ref{eq:ladder} using alternative GF representations is the topic of the following section. 


\section{Detailed Derivation of the Local Green’s Function in a Rectangular Cavity}
\label{sec:localGF}

This section completes the three–step singularity–subtraction procedure outlined in~\eqsref{eq:ladder}.  
Step~I removes the dipole singularity by subtracting the Green’s function of an infinite rectangular
waveguide.  
Step~II re-introduces the waveguide field in an alternative modal representation, which is then
regularised by subtracting the Green’s function of a parallel-plate waveguide.  
Step~III restores the parallel-plate contribution in still another representation and finally subtracts
the free-space Green’s function, yielding the desired \emph{local} dyadic Green’s function.

We first treat an electric dipole aligned with $\hat{\mathbf z}$; results for arbitrary orientation follow by rotation of coordinates.  
All quantities are time-harmonic with the $e^{j\omega t}$ convention and the cavity interior is vacuum.

Finally, for completeness and pedagogical clarity, we derive the subtracted Green’s dyad $\boldsymbol{G}^{s}(\boldsymbol{r},\boldsymbol{r}')$ 
and later obtain the local field by taking the coincidence limit
$\boldsymbol{r}\rightarrow\boldsymbol{r}'$.

\subsection{Local Electric and Magnetic Green’s Functions for a \texorpdfstring{$\hat{\mathbf z}$}{z}-Oriented Electric Dipole}
\label{subsec:GF_cavity}

\subsubsection{Step I: Cavity Minus Infinite Rectangular Waveguide}
\label{subsubsec:StepI}

\begin{figure}[!b]
  \centering
  \includegraphics[width=0.8\linewidth]{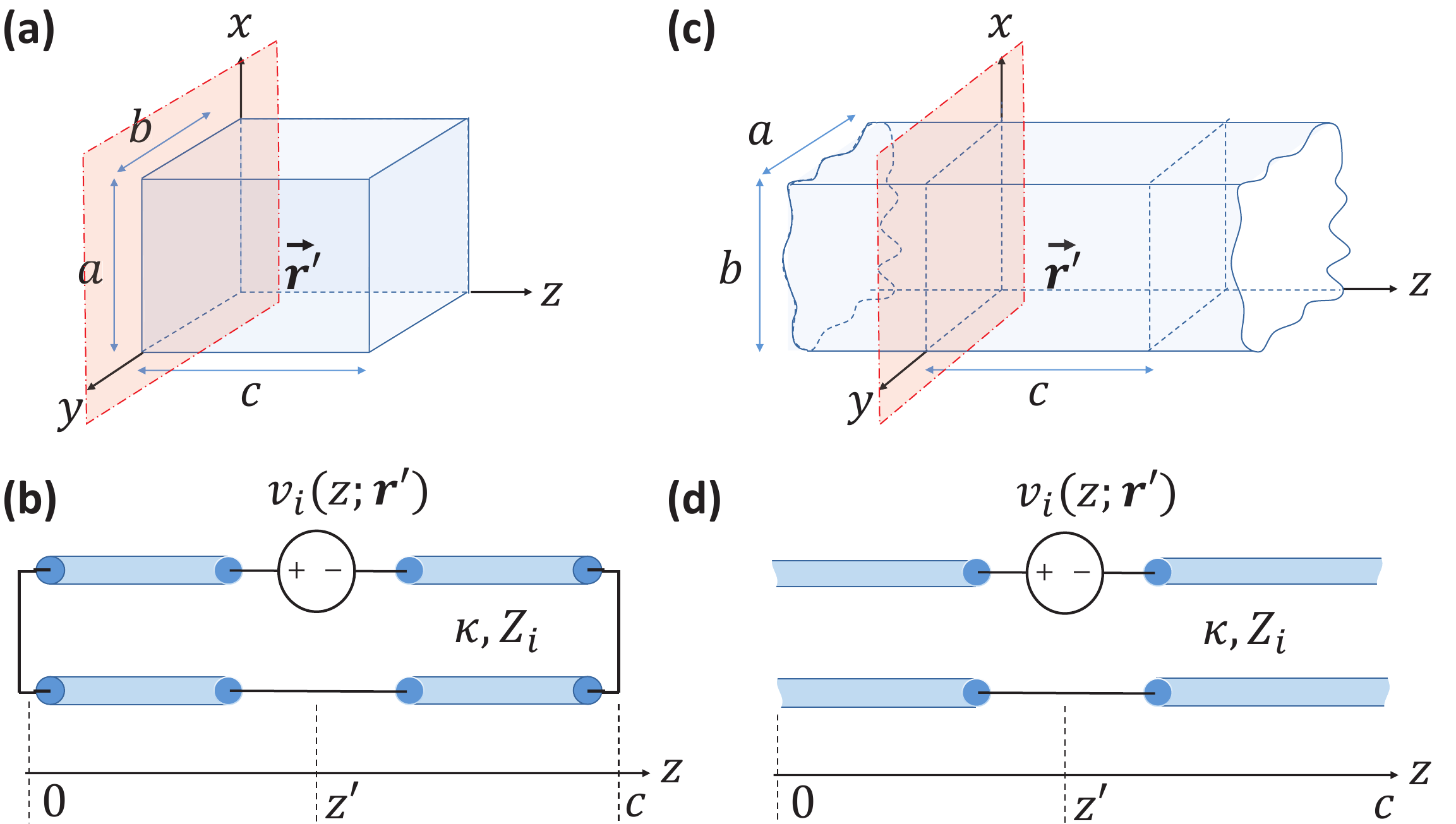}
  \caption{Step I subtraction.  
    (a) Rectangular cavity with $\hat{\mathbf z}$-polarised current dipole at $\mathbf r=\mathbf r'$.  
    (b) Equivalent 1-D transmission-line (TL) model (short-circuited).  
    (c) Infinite rectangular waveguide with identical source.  
    (d) Corresponding TL model (open at $\pm\infty$).}
  \label{fig:cavity}
\end{figure}

\paragraph*{Transverse eigenfunctions}
For the PEC cavity shown in Fig.~\ref{fig:cavity}(a) the 2-D Helmholtz problems on the $x-y$ plane
yield \cite{felsen2001radiation} (Sec. 3.2, Eqs.~(13) and (22) there)
\begin{subequations}
\label{eq:PhiPsi}
\begin{align}
  \Phi_{m,n}(x,y) &= \frac{2}{\sqrt{ab}}
  \sin\!\bigl(\tfrac{m\pi x}{a}\bigr)
  \sin\!\bigl(\tfrac{n\pi y}{b}\bigr),\!\! &&\text{(E-mode)}\label{eq:Phi}\\
  \Psi_{m,n}(x,y) &= \sqrt{\frac{\epsilon_m\epsilon_n}{ab}}
  \cos\!\bigl(\tfrac{m\pi x}{a}\bigr)
  \cos\!\bigl(\tfrac{n\pi y}{b}\bigr),\!\! &&\text{(H-mode)}\label{eq:Psi}
\end{align}
\end{subequations}
with $m,n\!\in\!\mathbb N$ and $\epsilon_m\!=\!2-\delta[m]$ where $\delta[m]$ denotes Kronecker's delta.

Using the basis functions in \eqsref{eq:PhiPsi}, the modal sources for E-and H-modes can be expanded for the physical domain sources 
$\mathbf J = J_0\delta(\mathbf r-\mathbf r')\hat{\mathbf z}$, $\mathbf M=\mathbf 0$ (in the following we take $J_0=1\mbox{[Am]}$ by definition of the dyadic Green's function, see Sec.~\ref{sec:math}).
By following this process, one readily observes that for this case only E-modes are excited; the modal sources are \cite{felsen2001radiation} (Sec. 2.4, Eq.~(21) there)
\begin{subequations}
\begin{align}
  &v_{m,n}(z;\mathbf r') =
    j\,\frac{k_{t,mn}}{\omega\varepsilon_0}\,
    \Phi_{m,n}(x',y')\,\delta(z-z'), \\ &
  i_{m,n}(z;\mathbf r') = 0,
\end{align}
\end{subequations}
with 
\begin{equation}
k_{t,mn}^2=(m\pi/a)^2+(n\pi/b)^2.
\end{equation}

By setting a unit voltage source in the corresponding TL models for the cavity and the RG we derive the modal voltages and currents 1D Green's function that, when multiplied by the modal sources, acts as the modal amplitudes of the transverse eigenfunctions that are required for the GF expansion. 
Specifically, for the associated infinite waveguide (Fig.~\ref{fig:cavity}(c,d)) 
\begin{subequations}
\label{eq:TL_RG1}
\begin{align}
  \mathcal V_{\mathrm{RG1},mn}(z,z') &= -\tfrac12\,\varsigma_{z,z'}e^{-j\kappa_{mn}|z-z'|},\\
  \mathcal I_{\mathrm{RG1},mn}(z,z') &= -\tfrac12 Y_{mn}'e^{-j\kappa_{mn}|z-z'|},
\end{align}
\end{subequations}
where $\varsigma_{z,z'}\!=\!\operatorname{sgn}(z-z')$,  $Y_{mn}'=1/Z'_{mn}=\omega\varepsilon_0/\kappa_{mn}$, and $\kappa_{mn}=\sqrt{k_0^2-k_{t,mn}^2}$ with $\mbox{Im}\{\kappa_{mn}\}>0$, $k_0=\omega\sqrt{\mu_0\varepsilon_0}$.

For the cavity configuration of Fig.~\ref{fig:cavity}(a,b) the PEC terminations at
\(z=0\) and \(z=c\) generate multiple reflections.  The resulting modal voltage
and current amplitudes for the \(mn\)-th E-mode read,

\begin{subequations}
\label{eq:VCV}
\begin{align}
\mathcal{V}_{{\mathrm{CV}},mn}(z,z') &=
  \frac{-1}{2\!\left(1-e^{-2j\kappa_{mn} c}\right)}
  \Bigl[
      \varsigma_{z,z'}\,e^{-j\kappa_{mn}|z-z'|}+\\ \notag
      e^{-j\kappa_{mn}(z+z')}  
      &-e^{-j\kappa_{mn}\!\left(2c-(z+z')\right)}
      -\varsigma_{z,z'}\,e^{-j\kappa_{mn}\!\left(2c-|z-z'|\right)}
  \Bigr],                                           \label{eq:VCV_V}\\[4pt]
\mathcal{I}_{{\mathrm{CV}},mn}(z,z') &=
  \frac{-Y_{mn}'}{2\!\left(1-e^{-2j\kappa_{mn} c}\right)}
  \Bigl[
      e^{-j\kappa_{mn}|z-z'|}+\\ \notag
      e^{-j\kappa_{mn}(z+z')} 
      &+e^{-j\kappa_{mn}\left(2c-(z+z')\right)}
      +e^{-j\kappa_{mn}\left(2c-|z-z'|\right)}
  \Bigr],                                           \label{eq:VCV_I}
\end{align}
\end{subequations}

\paragraph*{Subtracted fields}
Define
$\Delta\bar{\mathcal V}_{mn}^{(1)}\!=\!
  2(\mathcal V_{\mathrm{CV},mn}-\mathcal V_{\mathrm{RG1},mn})$ and
$\Delta\bar{\mathcal I}_{mn}^{(1)}\!=\!
  2Z_{mn}'(\mathcal I_{\mathrm{CV},mn}-\mathcal I_{\mathrm{RG1},mn})$.
With $S_x(t)\!=\!\sin(k_x t)$, $C_x(t)\!=\!\cos(k_x t)$, $S_y(t)\!=\!\sin(k_y t)$, $C_y(t)\!=\!\cos(k_y t)$
$k_x=m\pi/a$, $k_y=n\pi/b$, the Step I dyadic entries are \cite{felsen2001radiation} (Secs. 2.2, 2.3)
\begin{subequations}
\label{eq:StepI}
\begin{align}
  E_{xz}^{(1)} &=
    -\frac{2j}{\omega\varepsilon_0 ab}
    \sum_{m,n}\!
    \Delta\bar{\mathcal V}_{mn}^{(1)}(z,z')\,
    k_x S_x(x')S_y(y')C_x(x)S_y(y),\\
  E_{yz}^{(1)} &=
    -\frac{2j}{\omega\varepsilon_0 ab}
    \sum_{m,n}\!
    \Delta\bar{\mathcal V}_{mn}^{(1)}(z,z')\,
    k_y S_x(x')S_y(y')S_x(x)C_y(y),\\
  E_{zz}^{(1)} &=
     \frac{2}{\omega\varepsilon_0 ab}
     \sum_{m,n}\!
     \Delta\bar{\mathcal I}_{mn}^{(1)}(z,z')\,
     \frac{k_{t,mn}^{\,2}}{\kappa_{mn}}\times\,
    \notag\\  &\qquad \qquad \qquad \qquad S_x(x')S_y(y')S_x(x)S_y(y),\\
  H_{xz}^{(1)} &=
     \frac{2j}{ab}
     \sum_{m,n}\!
     \Delta\bar{\mathcal I}_{mn}^{(1)}(z,z')\,
     \frac{k_y}{\kappa_{mn}}\,
     S_x(x')S_y(y')S_x(x)C_y(y),\\
  H_{yz}^{(1)} &=
    -\frac{2j}{ab}
     \sum_{m,n}\!
     \Delta\bar{\mathcal I}_{mn}^{(1)}(z,z')\,
     \frac{k_x}{\kappa_{mn}}\,
     S_x(x')S_y(y')C_x(x)S_y(y),
\end{align}
\end{subequations}
where superscript~$(1)$ tracks Step I in~\eqsref{eq:ladder}.

\subsubsection{Step II: Alternative Waveguide Representation Minus Parallel-Plate Waveguide}
\label{subsubsec:StepII}

\begin{figure}[!t]
  \centering
  \includegraphics[width=0.8\linewidth]{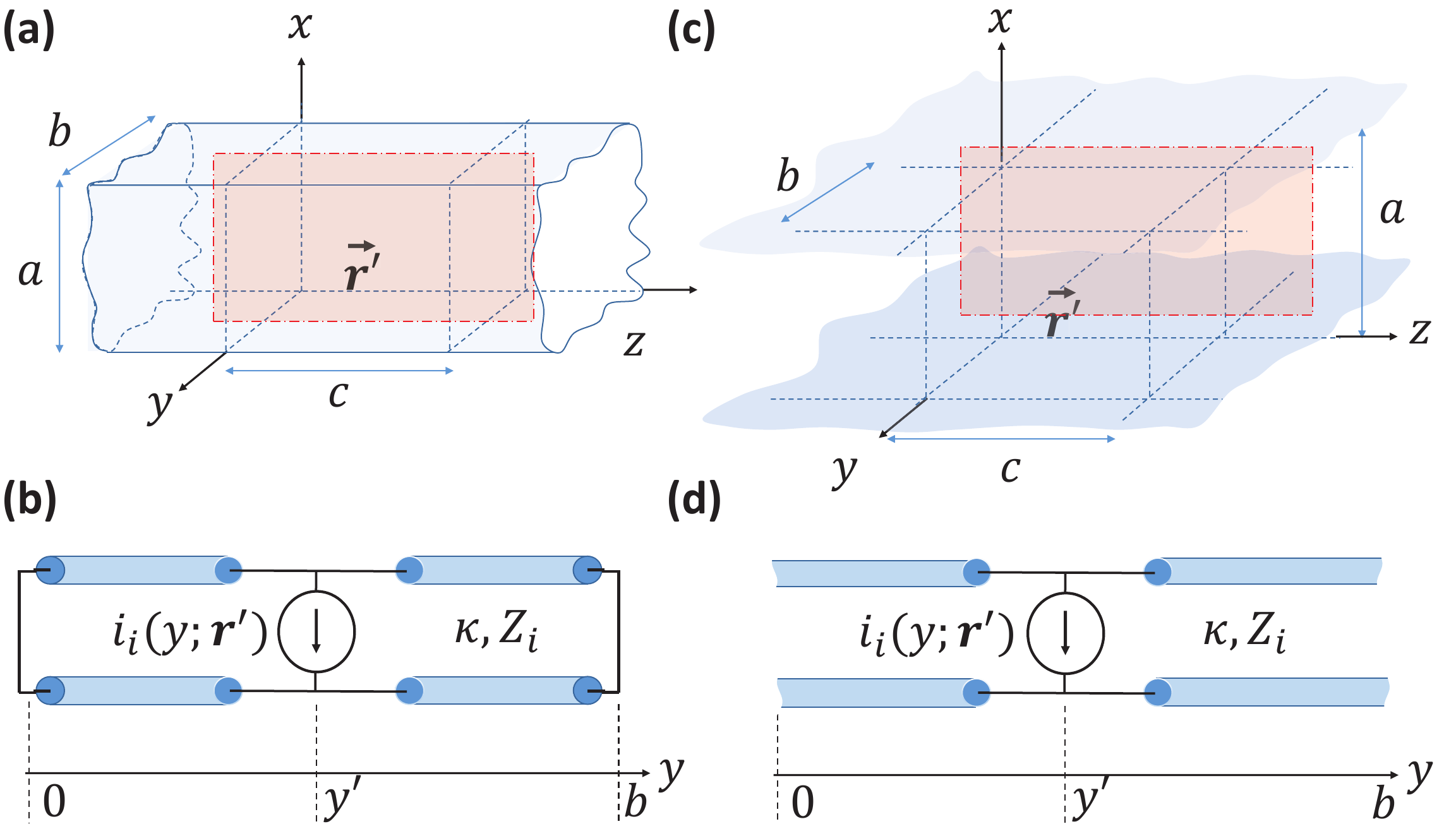}
  \caption{Step II subtraction.
    (a) Rectangular waveguide re-expanded with the $y$-axis as longitudinal direction (RG2).  
    (b) Corresponding TL model (short-circuited at $y=0,b$).  
    (c) Parallel-plate waveguide with the same source (PP1).  
    (d) Infinite TL model for PP1.}
  \label{fig:rect}
\end{figure}

\paragraph*{Alternative transverse plane}
Re-expand the waveguide field on the $x$–$z$ plane
so that $y$ is now the longitudinal variable (Fig.~\ref{fig:rect}(a)).
For each integer $m\!>\!0$ and continuous parameter $\xi\!\in\!\mathbb R$ \cite{felsen2001radiation}(Sec. 3.2, Eq.~(41) there),
\begin{subequations}
\label{eq:PhiPsi_RG2}
\begin{align}
  \Phi_{m,\xi}(x,z) &= \frac{1}{\sqrt{\pi a}}
    \sin\!\bigl(\tfrac{m\pi x}{a}\bigr)\,e^{-j\xi z}, &&\text{(E-mode)},\\
  \Psi_{m,\xi}(x,z) &= \sqrt{\frac{\epsilon_m}{2\pi a}}
    \cos\!\bigl(\tfrac{m\pi x}{a}\bigr)\,e^{-j\xi z}, &&\text{(H-mode)},
\end{align}
\end{subequations}
where $\epsilon_m$ is defined as above. Using the basis functions in \eqsref{eq:PhiPsi_RG2}, the modal sources for E-and H-modes can be expanded for the physical domain sources 
$\mathbf J = J_0\delta(\mathbf r-\mathbf r')\hat{\mathbf z}$, $\mathbf M=\mathbf 0$.
By following this process, one readily observes that for this case, both E- and H-modes are excited; For the E-modes the modal sources are \cite{felsen2001radiation}(Sec. 3.2)
\begin{subequations}
\label{eq:modal sources 2}
\begin{align}
   &i_i(y;\bm{r}')=-  \frac{j\xi}{k_{t,m\xi}} \Phi^*_{m,\xi}(x',z') \delta(y-y')
\\ &\mbox{and for H-modes,} \nonumber \\
   &i_i(y;\bm{r}')=  -\frac{1}{k_{t,m\xi}} \partial_x\Psi^{*}(x',z') \delta(y-y')
\end{align}
\end{subequations}
where superscript $*$ denotes complex conjugation, and with
\begin{equation}
k_{t,m\xi}^2=\xi^2+(m\pi/a)^2.
\end{equation}
By setting a unit current source in the TL models for the two problems shown in Fig.~\ref{fig:rect} we derive the 1D spectral Green's functions for this case.  
The RG is modelled by a short-circuited TL of length~$b$ (Fig.~\ref{fig:rect}(a,b)). Yielding
\begin{subequations}
\label{eq:VIRG2}
\begin{align}
  \mathcal V_{\mathrm{RG2},m\xi}(y,y') &=
    \frac{Z_{m\xi}}{2\!\bigl(e^{-2j\kappa_{m\xi} b}-1\bigr)}
    \Bigl[e^{-j\kappa_{m\xi}|y-y'|}-\\ \notag
          e^{-j\kappa_{m\xi}(y+y')} 
          &-e^{-j\kappa_{m\xi}\!\bigl(2b-(y+y')\bigr)}
          +e^{-j\kappa_{m\xi}\!\bigl(2b-|y-y'|\bigr)}\Bigr],\\
  \mathcal I_{\mathrm{RG2},m\xi}(y,y') &=
    \frac{1}{2\!\bigl(e^{-2j\kappa_{m\xi} b}-1\bigr)}
    \Bigl[\varsigma_{y,y'}e^{-j\kappa_{m\xi}|y-y'|}\\ \notag
          -e^{-j\kappa_{m\xi}(y+y')} 
          &+e^{-j\kappa_{m\xi}\!\bigl(2b-(y+y')\bigr)}
          -\varsigma_{y,y'}e^{-j\kappa_{m\xi}\!\bigl(2b-|y-y'|\bigr)}\Bigr].
\end{align}
\end{subequations}
with $Z_{m\xi}$ denoting $Z'_{m\xi}=\kappa_{m\xi}/\omega\varepsilon_0$ and $Z''_{m\xi}=\omega\mu_0/\kappa_{m\xi}$ for E- and H-modes, respectively, and where $\kappa_{m\xi}=\sqrt{k_0^2-k^2_{t,m\xi}},\quad \mbox{Im}\{\kappa_{m\xi}\}>0$.

Replacing the lateral PEC walls on $y=0,b$ transforms the RG waveguide to a parallel plate waveguide, and the corresponding TL model is unbounded as shown in Fig.~\ref{fig:rect}(c,d). This model yields,
\begin{subequations}
\label{eq:VIPP1}
\begin{align}
  &\mathcal V_{\mathrm{PP1}} =-\tfrac12 Z_{m,\xi}e^{-j\kappa|y-y'|},\\ &
  \mathcal I_{\mathrm{PP1}} =-\tfrac12 \varsigma_{y,y'}e^{-j\kappa|y-y'|}.
\end{align}
\end{subequations}

\paragraph*{Subtracted fields}
Define
$\Delta\bar{\mathcal V}^{(2)}_{m\xi}=2Z_{m,\xi}(\mathcal V_{\mathrm{RG2},m\xi}-\mathcal V_{\mathrm{PP1},m\xi})$,
$\Delta\bar{\mathcal I}^{(2)}_{m\xi}=2(\mathcal I_{\mathrm{RG2},m\xi}-I_{\mathrm{PP1},m\xi})$.
With $S_x(t),C_x(t)$ defined as before, the Step II entries are \cite{felsen2001radiation} (Secs.~2.2, 2.3)
\begin{subequations}
\label{eq:StepII}
\begin{align}
  E_{xz}^{(2)} =
    -\frac{j}{2\pi a\omega\varepsilon_0}
    &\sum_{m\neq0} S_x(x')C_x(x)\times\\ \notag
    &\int_{-\infty}^{\infty}\!
      \frac{\xi k_x}{\kappa}
      \Delta\bar{\mathcal{V}}^{(2)}_{m\xi}(y,y')\,
      e^{-j\xi(z-z')}\,d\xi,\\
  E_{yz}^{(2)} =
    -\frac{1}{2\pi a\omega\varepsilon_0}
    &\sum_{m\neq0} S_x(x')S_x(x)\times\\ \notag
    &
    \int_{-\infty}^{\infty}\!
      \xi\Delta\bar{\mathcal{I}}^{(2)}_{m\xi}(y,y')\,
      e^{-j\xi(z-z')}\,d\xi,\\
  E_{zz}^{(2)} =
     \frac{1}{2\pi a\omega\varepsilon_0}
     &\sum_{m\neq0} S_x(x')S_x(x)\times\\ \notag
    &
     \int_{-\infty}^{\infty}\!
      \frac{\kappa^2+k_x^2}{\kappa}
      \Delta\bar{\mathcal{V}}^{(2)}_{m\xi}(y,y')\,
      e^{-j\xi(z-z')}\,d\xi,\\
  H_{xz}^{(2)} =
     \frac{1}{2\pi a}
     &\sum_{m\neq0} S_x(x')S_x(x)\times\\ \notag
    &
     \int_{-\infty}^{\infty}\!
      \Delta\bar{\mathcal{I}}^{(2)}_{m\xi}(y,y')\,
      e^{-j\xi(z-z')}\,d\xi,\\
  H_{yz}^{(2)} =
    -\frac{j}{2\pi a}
     &\sum_{m\neq0} S_x(x')C_x(x)\times\\ \notag
    &
     \int_{-\infty}^{\infty}\!
      \frac{k_x}{\kappa}
      \Delta\bar{\mathcal{V}}^{(2)}_{m\xi}(y,y')\,
      e^{-j\xi(z-z')}\,d\xi,
\end{align}
\end{subequations}
with superscript~$(2)$ indicating Step II.

\subsubsection{Step III: Parallel-Plate Representation Minus Free Space}
\label{subsubsec:StepIII}

\begin{figure}[!b]
  \centering
\includegraphics[width=0.8\linewidth]{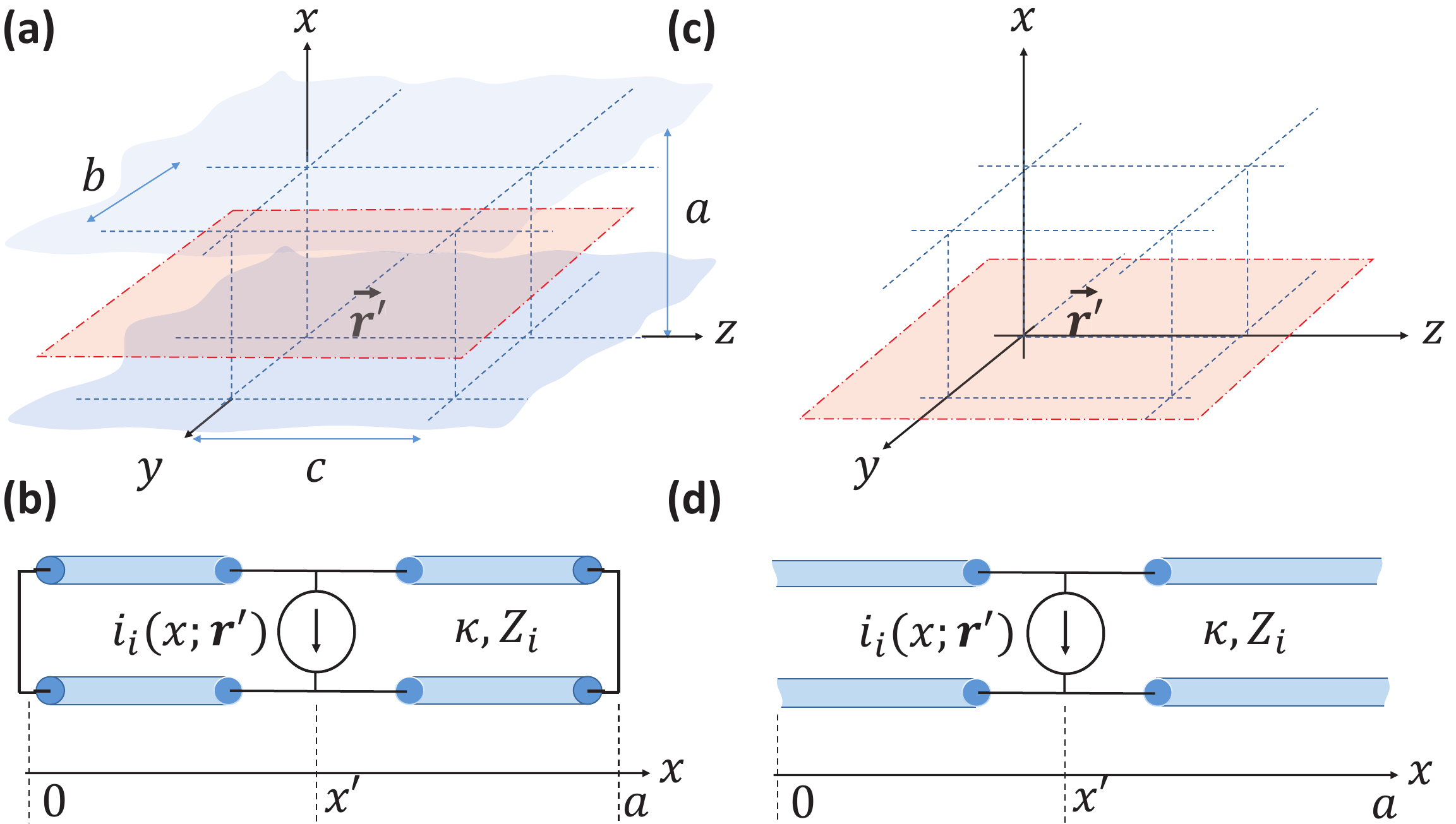}
  \caption{Step III subtraction.
    (a) Parallel-plate waveguide re-expanded with $x$ as longitudinal direction (PP2).  
    (b) Associated short-circuited TL.  
    (c) Free space with identical source.  
    (d) Infinite TL model for free space.}
  \label{fig:pp}
\end{figure}

Here we compensate for the previously subtracted PP Green's function and subtract the free space Green's function. Thus completing the third step in \eqsref{eq:ladder}. 
The configuration is shown in Fig.~(\ref{fig:pp}). The  $y-z$  plane is taken here as the transverse plane. The eigenfunctions set is continuous, and given by \cite{felsen2001radiation} (Sec. 3.2, Eq.~(40) there)
\begin{equation}\label{eq:modal fs}
\Phi_{\xi\eta}(y,z)=\Psi_{\xi\eta}(y,z)=\frac{1}{2\pi} e^{-j\eta y} e^{-j\xi z}.
\end{equation}
The modal sources for E- and H- modes are solely currents, and given by \eqsref{eq:modal sources 2} after a trivial change of variables, $a\leftrightarrow b, x\leftrightarrow y$. 
The modal 1D GF are given by ~\eqsref{eq:VIRG2} after similar variable change, For the PP2 model
(Fig.~\ref{fig:pp}(a,b)):
\begin{align}
  \mathcal V_{\mathrm{PP2}}(x,x') &= \mathcal V_{\mathrm{RG2}}(y,y')\Big|_{x\leftrightarrow y,\;a\leftrightarrow b},\\
  \mathcal I_{\mathrm{PP2}}(x,x') &= \mathcal I_{\mathrm{RG2}}(y,y')\Big|_{x\leftrightarrow y,\;a\leftrightarrow b}.
\end{align}
Likewise, for the free space model (Fig.~\ref{fig:pp}(c,d)) we use \eqsref{eq:VIPP1},
\begin{subequations}
\begin{align}
  &\mathcal V_{\mathrm{FS}}(x,x') = \mathcal V_{\mathrm{PP1}}(y,y')\Big|_{x\leftrightarrow y,\;a\leftrightarrow b},\\ &
  \mathcal I_{\mathrm{FS}}(x,x') = \mathcal I_{\mathrm{PP1}}(y,y')\Big|_{x\leftrightarrow y,\;a\leftrightarrow b}.
\end{align}
\end{subequations}
Here, 
\begin{equation}
k^2_{t,\xi\eta} = \xi^2+\eta^2, \kappa_{\xi\eta} = \sqrt{k_0^2-k^2_{t,\xi\eta}}, \mbox{Im}\{\kappa_{\xi\eta}\}>0
\end{equation}
\paragraph*{Subtracted fields}
Define
$\bar{\mathcal V}_{\xi\eta}^{(3)}=2Z_{\xi\eta}(\mathcal V_{\mathrm{PP2},\xi\eta}-\mathcal V_{\mathrm{FS},\xi\eta})$,
$\bar{\mathcal I}_{\xi\eta}^{(3)}=2(\mathcal I_{\mathrm{PP2},\xi\eta}-\mathcal I_{\mathrm{FS},\xi\eta})$.
Let $(\bar\rho,\bar\phi)$ be polar coordinates of $(z-z',y-y')$ (i.e., $z-z'=\bar\rho\cos\bar\phi$, and $y-y'=\bar\rho\sin\bar\phi$ ) and set for brevity of notations 
$q^2\!=\!k_{t,\xi\eta}^2$.
The subtracted fields due to Step III in \eqsref{eq:ladder} are \cite{felsen2001radiation} (Secs.~ 2.3, 2.4)
\begin{subequations}
\label{eq:StepIII}
\begin{align}
  E_{xz}^{(3)} &=
    \frac{j\cos\bar\phi}{4\pi\omega\varepsilon_0}
    \int_{0}^{\infty}\!
      q^{2}\mathcal J_{1}(q\bar\rho)
\bar{\mathcal{I}}_{\xi\eta}^{(3)}(x,x')\,
      dq,\\
  E_{yz}^{(3)} &=
    \frac{\sin(2\bar\phi)}{8\pi\omega\varepsilon_0}
    \int_{0}^{\infty}\!
      \frac{q^{3}}{\kappa_{\xi\eta}}\mathcal J_{2}(q\bar\rho)
\bar{\mathcal{V}}_{\xi\eta}^{(3)}(x,x')\,
      dq,\\
  E_{zz}^{(3)} &=
    \frac{1}{4\pi\omega\varepsilon_0}
    \int_{0}^{\infty}\!
      \!\Bigl[\!
        \tfrac{q^{3}}{2\kappa_{\xi\eta}}\cos(2\bar\phi)\mathcal J_{2}(q\bar\rho)
        +\\ \notag &\qquad q\!\bigl(2k^{2}-q^{2}\bigr)\!\tfrac{\mathcal J_{0}(q\bar\rho)}{\kappa_{\xi\eta}}
      \Bigr]   \!\bar{\mathcal{V}}_{\xi\eta}^{(3)}(x,x')\,
      dq,\\
  H_{xz}^{(3)} &=
    -\frac{j\sin\bar\phi}{4\pi}
    \int_{0}^{\infty}\!
      \frac{q^{2}}{\kappa_{\xi\eta}}\mathcal J_{1}(q\bar\rho)   \bar{\mathcal{V}}_{\xi\eta}^{(3)}(x,x')\,
      dq,\\
  H_{yz}^{(3)} &=
    -\frac{1}{4\pi}
    \int_{0}^{\infty}\!
      q\,\mathcal J_{0}(q\bar\rho)
      \bar{\mathcal{I}}_{\xi\eta}^{(3)}(x,x')\,
      dq,
\end{align}
\end{subequations}
where $\mathcal{J}_n(t)$ represents the bessel function of the first kind and order $n$, ans with superscript~$(3)$ marking Step III in \eqsref{eq:ladder}.

\subsubsection{Assembly of the Local Dyadic Green’s Function}

Following eqref{eq:ladder}, adding the three regularised contributions gives, for example, the $xz$-component of the secondary dyadic GF:
\begin{subequations}
\label{eq:secondary xz}
\begin{align}
  \mathbf G^{s,ee}_{xz}(\mathbf r,\mathbf r') &=
    \hat{\mathbf x}\hat{\mathbf z}\,
    \bigl(E_{xz}^{(1)}+E_{xz}^{(2)}+E_{xz}^{(3)}\bigr),\\
  \mathbf G^{s,he}_{xz}(\mathbf r,\mathbf r') &=
    \hat{\mathbf x}\hat{\mathbf z}\,
    \bigl(H_{xz}^{(1)}+H_{xz}^{(2)}+H_{xz}^{(3)}\bigr).
\end{align}
\end{subequations}
All other dyadic components follow analogously. Then, by \eqsref{eq:sec to local} the local field Green's function is obtained by setting $\mathbf{r}=\mathbf{r}'$ in \eqsref{eq:secondary xz}. 


\section{Stability and Spectral Efficiency of the Subtraction Scheme}
\label{sec:stability_efficiency}

Direct evaluation of the local field in~\eqsref{eq:sec to local}
demands a numerical limit of the form ``$\infty-\infty$,'' an
operation that is notoriously ill-conditioned.
Our remedy is to carry out \emph{subtractions in the spectral
domain} rather than in real space.
In this way the cancellation of two \(3\text{-D}\) singular fields
(\(\propto 1/r^{3}\) as \(r\!\to\!0\))
is replaced by the cancellation of two
\(1\text{-D}\) Green’s functions (GFs) whose only singularity is a
finite jump.
Crucially, each difference is taken between GFs \emph{expanded in the
same spectral basis}, hence the ladder representation
in~\eqsref{eq:ladder} that employs the alternative GF forms
summarised in Sec.~\ref{sec:localGF}.
After the trivial source-term extraction, every residual kernel
\(\Delta\bar{\mathcal V}^{(i)}\) and
\(\Delta\bar{\mathcal I}^{(i)}\) \((i=1,2,3)\)
decays exponentially with its spectral variable—even when
\(\mathbf r=\mathbf r'\).
Consequently, each term of the ladder in~\eqsref{eq:ladder} converges
exponentially fast.

\begin{figure}[!t]
  \centering
  \subfloat[Step~I kernel
            \(\Delta\bar{\mathcal V}^{(1)}_{mn}(z',z')\)
            versus mode index~\(m\) for \(n=0,4\).]%
           {\includegraphics[width=0.48\linewidth]{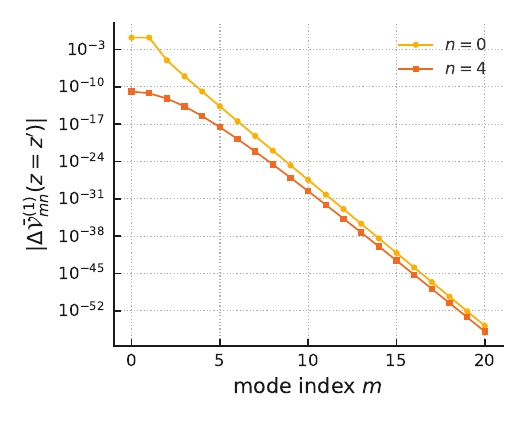}}%
  \hfill
  \subfloat[Step~II kernel
            \(\Delta\bar{\mathcal V}^{(2)}_{m\xi}(y',y')\)
            versus \(\xi/k_{0}\) for \(m=0,4\).]%
           {\includegraphics[width=0.48\linewidth]{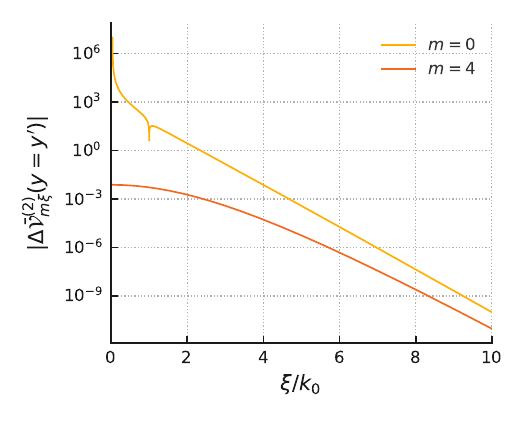}}
  \caption{Spectral kernels after source-singularity subtraction.
           Both panels reveal the exponential decay that underpins
           the spectral efficiency of the proposed scheme.}
  \label{fig:conv}
\end{figure}

We illustrate the point for the component
\(E_{xz}^{(1,2,3)}\).
Figure~\ref{fig:conv} compares the spectral kernels that appear in
Steps~I–II:
\begin{equation*}
\Delta\bar{\mathcal V}^{(1)}_{mn}(z',z')
\quad\text{and}\quad
\Delta\bar{\mathcal V}^{(2)}_{m\xi}(y',y').
\end{equation*}
Panel~(a) plots \(\Delta\bar{\mathcal V}^{(1)}_{mn}\)
versus~\(m\) for \(n=0\) and \(n=4\);
panel~(b) plots
\(\Delta\bar{\mathcal V}^{(2)}_{m\xi}\)
versus the continuous parameter \(\xi\in[0,10k_{0}]\) for
\(m=0\) and \(m=4\).
In both cases the exponential envelope is unmistakable,
guaranteeing rapid numerical convergence of the associated ladder
entries.
For Step~III the same exponential decay in $\bar{\mathcal{V}}^{(3)}$ exists; however, in this case, the kernel vanishes identically because
\(\mathcal J_{1}(0)=0\), hence \(E_{xz}^{(3)}=0\).

\section{Dyadic Green Functions for Arbitrary Source Polarization}
\label{sec:level_gen}

Sections~\ref{sec:localGF} and Appendix~\ref{sec:level_3b} give the
local Green functions for electric and magnetic
dipoles oriented along~\(\hat z\) inside a rectangular cavity.
We extend those results here to an \emph{arbitrarily polarized} dipole
by simple axis permutations.

Let the electric and magnetic current densities be
\(\mathbf J=(J_{x},J_{y},J_{z})\) and
\(\mathbf M=(M_{x},M_{y},M_{z})\),
respectively, inside a cavity of size \(a\times b\times c\).
Dyadic components for \(\hat z\)-sources,
\(G^{f}_{z'z'}\), \(G^{f}_{x'z'}\), and \(G^{f}_{y'z'}\),
with \(f\!\in\!\{ee,\,em,\,me,\,mm\}\),
have already been derived.
The remaining components follow by cyclic coordinate swaps:
\vspace{2pt}
\subsubsection*{Dipole along \(\hat x\)}
\vspace{-4pt}
\begin{enumerate}
  \item Rotate axes as
        \((x,y,z)\!\mapsto\!(z',x',y')\).
  \item Relabel cavity dimensions
        \((a,b,c)\!\mapsto\!(c',a',b')\).
  \item Substitute
        \(
        G^{f}_{xx}\!\leftrightarrow\!G^{f}_{z'z'},\;
        G^{f}_{yx}\!\leftrightarrow\!G^{f}_{x'z'},\;
        G^{f}_{zx}\!\leftrightarrow\!G^{f}_{y'z'}.
        \)
\end{enumerate}
\vspace{2pt}
\subsubsection*{Dipole along \(\hat y\)}
\vspace{-4pt}
\begin{enumerate}
  \item Rotate axes as
        \((x,y,z)\!\mapsto\!(y',z',x')\).
  \item Relabel dimensions
        \((a,b,c)\!\mapsto\!(b',c',a')\).
  \item Substitute
        \(
        G^{f}_{xy}\!\leftrightarrow\!G^{f}_{y'z'},\;
        G^{f}_{yy}\!\leftrightarrow\!G^{f}_{z'z'},\;
        G^{f}_{zy}\!\leftrightarrow\!G^{f}_{x'z'}.
        \)
\end{enumerate}

\vspace{2pt}
Applying these permutations to every
\(f\in\{ee,\,em,\,me,\,mm\}\)
yields the complete electric–magnetic dyadic
for any dipole orientation.
The explicit expressions, too lengthy for print, are fully coded and
available upon reasonable request.

\section*{Why Not Just Use a Commercial Solver?}\label{Sec:Why not Full Wave}

A natural question arises: can the joint cavity--particle resonances be computed more easily using commercial full-wave eigenmode solvers such as COMSOL, CST, or HFSS?

In principle, these tools can solve Maxwell's equations in arbitrary geometries, even in the presence of dispersive materials. However, in practice, they face severe limitations in scenarios similar to the one addressed here, namely, a dispersive subwavelength resonator embedded in a closed cavity with dense spectral features and hybrid modes \cite{Lalanne2019QNM,RamanFan2010Hermitian,Xiao2021FEMDispersion,Xiao2025Holomorphic,Tisseur2020RieszNEP}
. We briefly highlight several major challenges:

\begin{itemize}[label={},leftmargin=0pt,labelsep=0.5em]
\item \textbf{Nonlinear eigenvalue problem:} In dispersive media (e.g., Drude or Lorentz models), the permittivity $\varepsilon(\omega)$ depends on the eigenfrequency itself. This turns the linear eigenvalue problem into a nonlinear one. Standard solvers often linearize the problem by fixing frequency-dependent parameters at a guessed value, introducing significant errors or convergence failures unless a sophisticated Newton-type root-tracking algorithm is employed.

\item \textbf{Mode crowding and hybridization:} In strongly coupled systems like the one considered here, cavity modes and localized particle resonances can lie within a few percent of each other, producing multiple closely spaced or even nearly degenerate eigenmodes. Commercial solvers may converge to one solution, miss others, or yield inaccurate results without manual mode tracking and multiple initialization points.

\item \textbf{High computational cost:} Achieving convergence with fine spectral resolution near hybridization points requires extremely fine frequency sweeps, dense meshing, and manual solver tuning. In contrast, the proposed semi-analytical approach exploits modal decomposition and Green's function hierarchy to achieve \textit{exponential convergence} with minimal computational effort. In a manner akin to the Lindblad master equation \cite{Manzano}, in our formulation, the entire environment is \emph{exactly} encapsulated within a single complex dyad -- the renormalized local field GF. Thus, leading to a simple, numerically convenient algebraic equation for the eigenfrequencies \eqsref{eq:eig}.
\end{itemize}

Furthermore, the physics of local field renormalization---specifically the precise cancellation of radiation corrections and the enforcement of energy conservation---is difficult to observe or validate within a black-box eigenmode solver. Our formulation provides not only accurate eigenfrequencies, but also detailed physical insight into the origin and nature of each mode, which is essential in strongly coupled and chiral scenarios.
For these reasons, while commercial solvers are useful for many classical electromagnetic problems, they are not ideally suited for the rigorous evaluation of cavity–particle eigenmodes in the deep-subwavelength, dispersive, and strongly coupled regimes considered here. This highlights the necessity of the present formalism.


\section{Numerical Examples}
\label{sec:level_ex}

We apply the local‐field formulation to a single plasmonic sphere embedded in a rectangular cavity. Three material models are considered:

\begin{enumerate}
  \item isotropic permittivity (Drude plasma);
  \item gyrotropic (magnetised) Drude plasma;
  \item bi‐isotropic (chiral) medium.
\end{enumerate}

The analysis is fully self-consistent—no weak-coupling or mode-truncation assumptions are introduced.

The cavity dimensions are
\(a=b=10~\mu\text{m},\;c=30~\mu\text{m}\);
its first three cut-off frequencies are
\(\omega_{1,\mathrm{cv}}\!\approx\!15.81~\text{THz},\;
 \omega_{2,\mathrm{cv}}\!\approx\!18.03~\text{THz},\;
 \omega_{3,\mathrm{cv}}\!\approx\!21.21~\text{THz}\).
The sphere is located at $\bm{r}'=(0.2a,0.66b,0.365c)$ and its radius is \(R_{0}=1~\mu\text{m}\), so
\(2R_{0}\!\ll\!\lambda\) at the lowest cavity resonance and
\(2R_{0}\!\ll\!a,b,c\); the electric–dipole approximation is therefore
valid in all cases. 
Full wave numerical validation using CST and nonlinear eigenvalue extraction algorithm is provided just for the simplest case of isotropic Drude plasma. For the gyrotropic and chirala partticles using full wave solver becomes impractical. 

\subsection{Isotropic Plasmonic Sphere}
\label{isotropic}

\subsubsection*{Formulation}

A lossless Drude sphere of radius \(R_{0}=1~\mu\text{m}\) is centred in
the cavity.  Within the discrete–dipole approximation its (dynamic)
electric polarizability is
\begin{subequations}
\begin{align}
\boldsymbol{\alpha}^{-1}
      &=\Bigl[\frac{1}{\varepsilon_{0}V}
              \frac{\varepsilon_{r}(\omega)+2}
                   {3\,[\varepsilon_{r}(\omega)-1]}
              +j\frac{k^{3}}{6\pi\varepsilon_{0}}\Bigr]\,
        \mathbf{I}_3,
\\
\varepsilon_{r}(\omega) &= 1-\frac{\omega_{p}^{2}}{\omega^{2}},
\end{align}
\end{subequations}
where \(\omega_{p}\) is the plasma frequency and
\(V=4\pi R_{0}^{3}/3\).
Resonances of the coupled sphere–cavity system satisfy
\eqsref{eq:eig}.

\subsubsection*{Numerical Results}

\begin{figure}[!b]
  \centering
  \includegraphics[width=\linewidth]{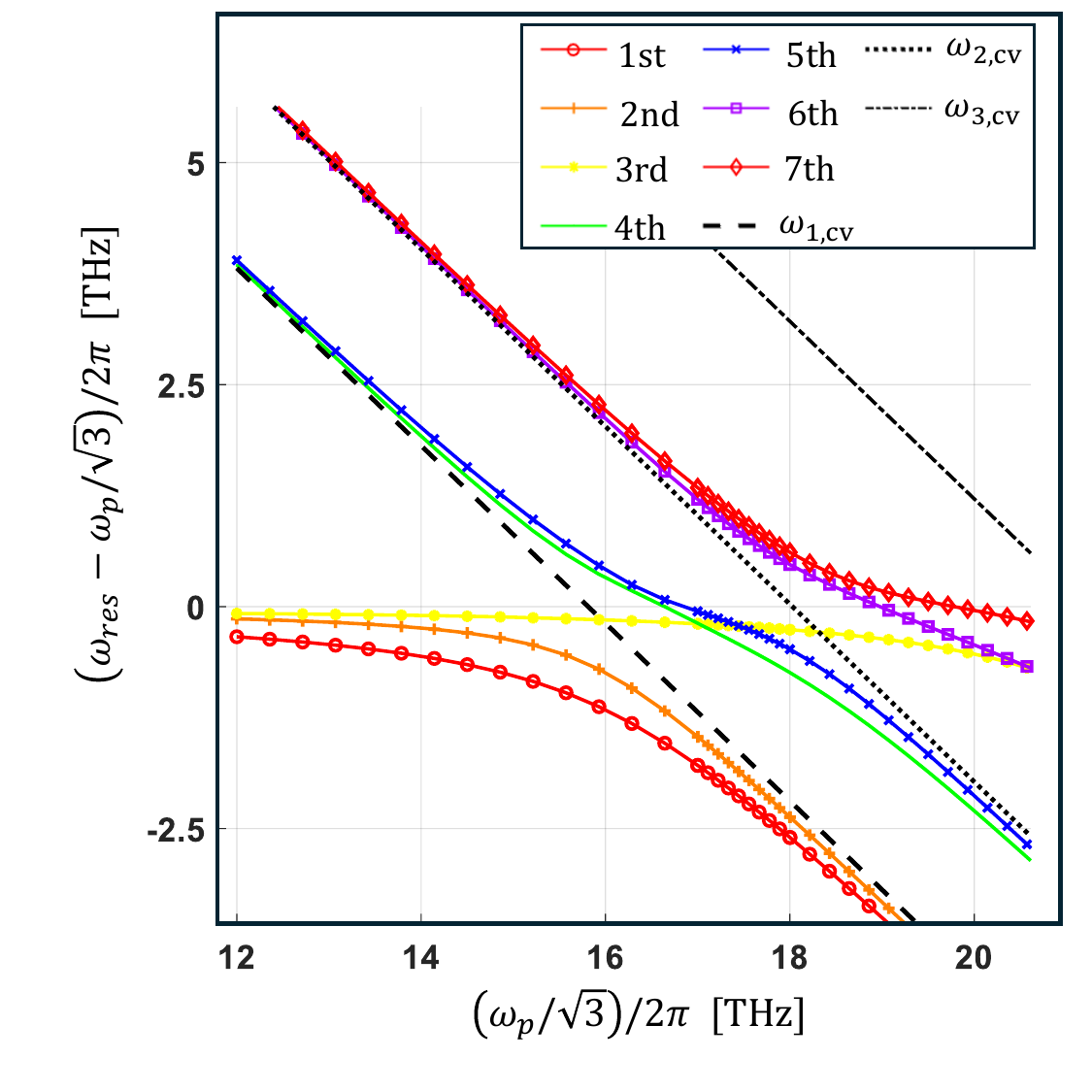}
  \caption{Evolution of the seven lowest coupled resonances versus the
           isolated‐particle frequency
           \(\omega_{p}/\sqrt{3}\).
           Dashed lines mark the first three cavity cut‐offs
           \(\omega_{i,\mathrm{cv}}\).}
  \label{fig:epsart_summ}
\end{figure}

Figure~\ref{fig:epsart_summ} traces the lowest resonances as
\(\omega_{p}\) is swept up to the third empty‐cavity cut-off.
Because \(a=b\), each empty cavity mode is doubly degenerate.
Two resonance families emerge:
\begin{itemize}
\item \textbf{Particle–dominated} (\emph{cavity/particle}):
      modes~1–3 originate below the first cut‐off.
      Modes~1–2 are excited by TE fields; mode~3 by a \(z\)-polarised
      field.
      Increasing \(\omega_{p}\) pulls modes~1–2 toward the first
      cut-off, whereas mode~3 remains almost fixed.

\item \textbf{Cavity–dominated} (\emph{particle/cavity}):
      modes~4–7 start near a cavity resonance and drift toward the
      next as \(\omega_{p}\) grows—behaviour beyond perturbative
      models.
\end{itemize}

Coupling therefore lifts all degeneracies and produces a rich,
non-perturbative rearrangement of the spectrum.

\subsubsection*{A note on energy conservation}
Because the cavity and the particle are lossless, every physical resonance must occur at a \emph{purely real} frequency. Individually, however, both the dynamic polarizability and the local‐field Green dyadic are \emph{complex}. A real root can therefore emerge only if their imaginary parts cancel exactly. The fact that our numerical sweeps always return real-valued resonances confirms that the proposed regularisation scheme removes the divergent self-terms with high fidelity.

\subsection{Isotropic Plasmonic Sphere — Numerical Extraction of Eigenfrequencies}

As discussed in Sec.~\ref{Sec:Why not Full Wave}, incorporating material dispersion into the eigenvalue problem transforms it into a nonlinear eigenvalue problem. This presents a significant challenge, particularly in the presence of multiple closely spaced resonances, where standard eigenmode solvers may struggle to converge or fail to capture all solutions.

Nevertheless, in the simplest case—a lossless, isotropic Drude model—it is still possible to devise a tractable numerical scheme using the so-called \emph{auxiliary eigenvalue problem} (AEP) or \emph{parameter mapping technique}. 
Instead of directly solving the inherently nonlinear eigenvalue problem—where material parameters such as permittivity depend explicitly on the eigenfrequency—one first constructs and solves a family of linear eigenvalue problems by treating the material parameter (e.g., \( \epsilon_{\text{sph}} \)) as a fixed, dispersionless value. The corresponding eigenfrequencies are extracted for each value in the parameter sweep. In the second stage, a known analytical relation (e.g., the Drude model) is used to map each swept parameter value back to its associated physical resonance frequency. This two-step strategy offers a tractable and numerically feasible route for tackling nonlinear eigenvalue problems that arise in simple dispersive systems. However, as the number of material parameters increases, e.g., when the material is anisotropic or chiral, the parametric sweep over all the entries of the permittivity dyad becomes intractable. 

\begin{figure}[t]
  \centering
  \hspace{-1cm}
  \includegraphics[width=\columnwidth]{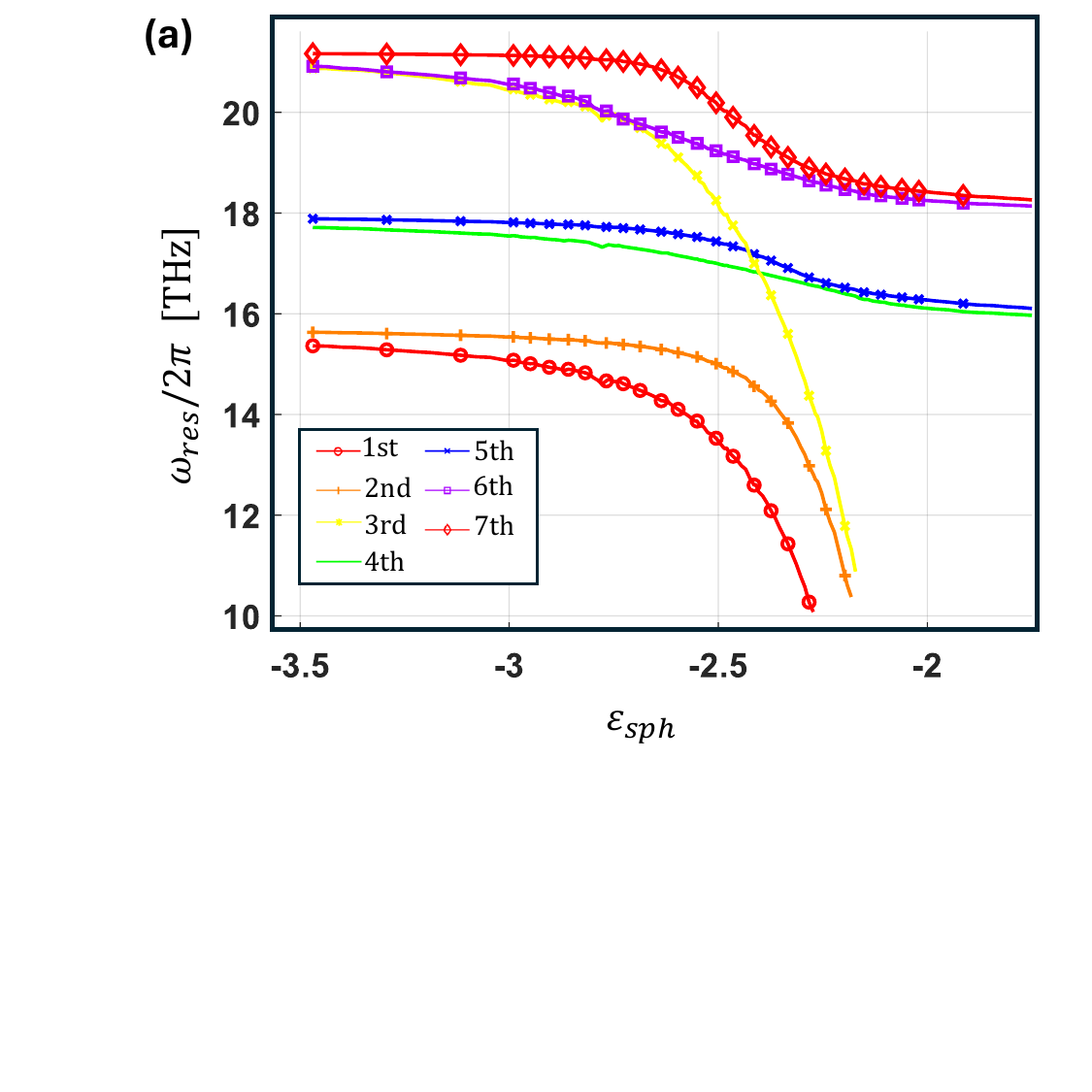} 
  \vspace{-2.5cm}
  
  \includegraphics[width=\columnwidth]{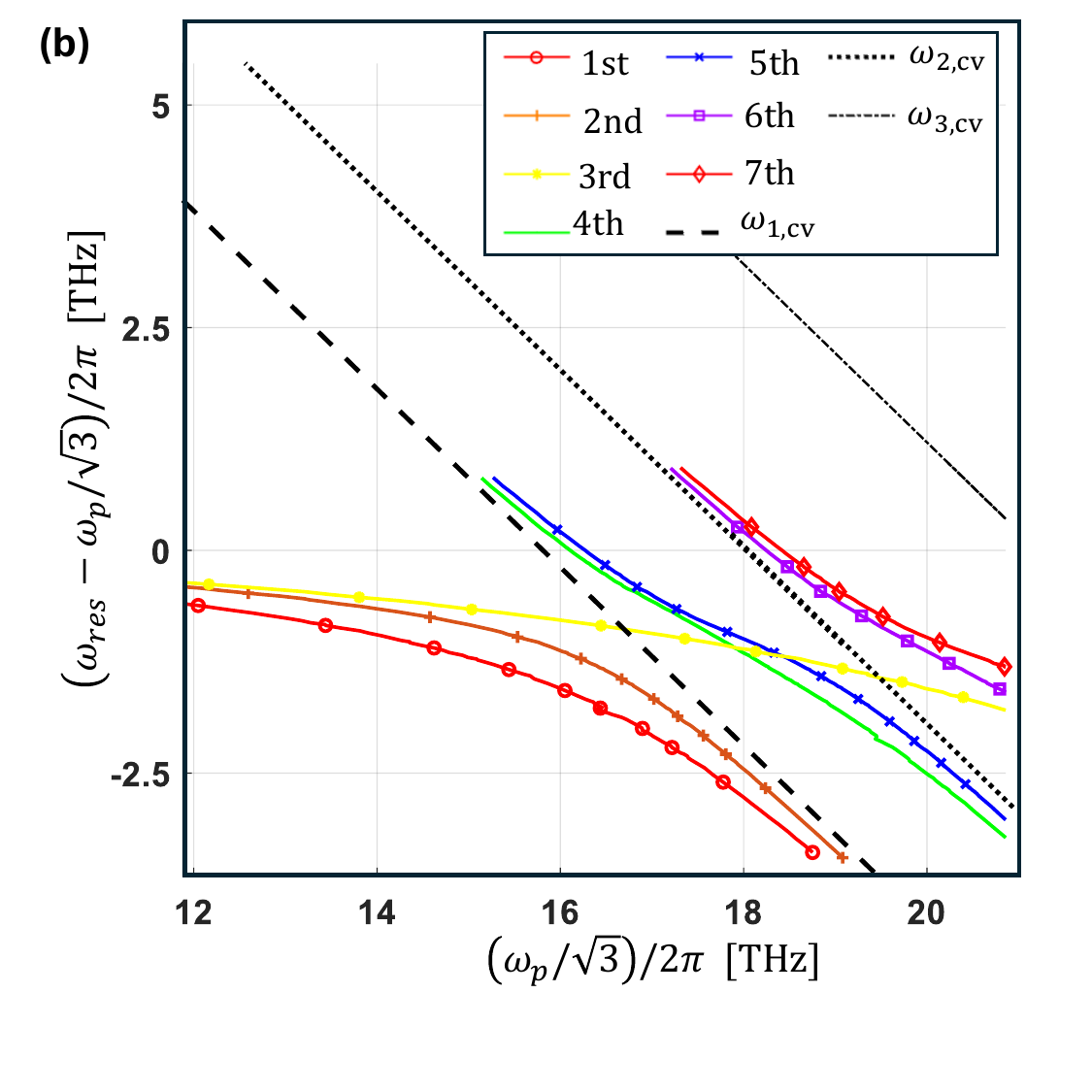}

   \vspace{-1cm}
  \caption{Numerical results extracted from CST eigenmode solver and mapped to a dispersive Drude model.  
  (a)~Ordered and interpolated eigenvalue raw data from CST Studio.  
  (b)~Dispersion of joint resonance frequencies versus \( \omega_p \).}
  \label{fig:numerical_combined}
\end{figure}

 Here, we employ a commercial solver (CST Studio Suite) to compute the eigenfrequencies of a spherical plasmonic particle embedded in a rectangular cavity, thereby enabling direct comparison with the results of Sec.~\ref{isotropic}. To this end, we first assume the particle is characterized by a constant negative permittivity \( \epsilon_{\text{sph}} \), i.e., treated as \emph{dispersionless} during the solver run, thereby rendering the eigenvalue problem linear. A parametric sweep over \( \epsilon_{\text{sph}} \) is then performed, with CST used to extract the corresponding eigenfrequencies \( \omega_{\text{res},i} \) at each step.
The computed eigenfrequencies, comprising seven distinct modes, are presented in Fig.~\ref{fig:numerical_combined}(a).

In the second stage, \emph{to connect these numerically extracted eigenfrequencies to a physical dispersive model}, we apply the standard lossless Drude relation,
\(
\epsilon_{\text{sph}} = 1 - {\omega_p^2}/{\omega^2},
\)
to construct a nonlinear mapping between the sweep parameter \( \epsilon_{\text{sph}} \) used in the solver and the physical frequency \( \omega \). Substituting each eigenfrequency \( \omega = \omega_{\text{res},i} \) into the expression yields a nonlinear equation relating the plasma frequency \( \omega_p \) to the resonance frequency \( \omega_{\text{res},i} \). Solving this equation provides \( \omega_{\text{res},i}(\omega_p) \), which is plotted in Fig.~\ref{fig:numerical_combined}(b).

It is worth noting that in our simulation, the particle is positioned adjacent to the cavity wall (with  \( x'=0.2a = 2\,\mu\text{m} \) and cavity radius \( R_0 = 1\,\mu\text{m} \)), placing the system beyond the formal validity of the discrete dipole approximation and polarizability-based theories \cite{Tretyakov2011}. Nevertheless, the CST full-wave results, combined with the described numerical procedure, exhibit both qualitative and quantitative agreement with our regularized semi-analytical predictions shown in Fig.~\ref{fig:epsart_summ}.

While this numerical strategy is effective for the idealized Drude model, it becomes prohibitively complex for more realistic scenarios. In particular, extending the method to lossy, multi-resonant, gyrotropic, or chiral media is generally infeasible due to the lack of an explicit and invertible mapping between \( \epsilon_{\text{sph}} \) and the underlying material parameters. This severely limits the viability of parametric sweeps and post-processing in such cases.

\subsection{Gyrotropic Sphere Inside the Cavity}
\label{subsec:gyro}

\subsubsection*{Problem Formulation}

A plasmonic Drude sphere of volume \(V\) is placed at the cavity
centre and biased by a static magnetic field
\(\mathbf B_{0}=B_{0}\hat z\).
Because \(R_{0}\ll a,b,c\), the dipole approximation remains valid.
The dynamic \emph{electric} polarizability—including radiation
damping—reads\cite{ishimaru1991electromagnetic,Sadzi2024b}
\begin{subequations}            
\begin{align}
\boldsymbol{\alpha}^{-1}
  &= \frac{k^{3}}{6\pi\varepsilon_{0}}
     \bigl(\boldsymbol{\alpha}_{h}^{-1}+j\mathbf I_{3}\bigr),\\
\boldsymbol{\alpha}_{h}^{-1}
  &= \begin{bmatrix}
       g_{xx} & -j g_{xy} & 0\\
       j g_{xy} & g_{yy} & 0\\
       0 & 0 & g_{zz}
     \end{bmatrix}.
\end{align}
\end{subequations}%
with
\begin{subequations}
\begin{align}
g_{uu}&=\frac{6\pi}{k^{3}V}
        \!\left[\frac13
          -\frac{\omega(\omega-j\gamma)}
                 {\omega_{p}^{2}}
           \frac{(\omega-j\gamma)^{2}-\omega_{c}^{2}}
                {(\omega-j\gamma)^{2}+\omega_{c}^{2}}\right],  
                \\
g_{zz}&=\frac{6\pi}{k^{3}V}
        \!\left[\frac13
          -\frac{\omega(\omega-j\gamma)}{\omega_{p}^{2}}\right],       
          \\
g_{xy}&=\frac{6\pi}{k^{3}V}
        \!
        \frac{(\omega-j\gamma)\omega_{c}}{\omega_{p}^{2}} \times \notag
               \\
       &\left[\frac{(\omega-j\gamma)^{2}-\omega_{c}^{2}}
                     {(\omega-j\gamma)^{2}+\omega_{c}^{2}}
            -\frac23 \frac{\omega_{c}(\omega-j\gamma)}
             {(\omega-j\gamma)^{2}+\omega_{c}^{2}} ] 
                     \right]
\end{align}
\end{subequations}
where \(\omega_{c}=|q_{e}|B_{0}/m_{e}\) is the cyclotron frequency, and the damping rate is set to $\gamma=0$.

\paragraph*{Isolated–particle resonances.}
The roots of
\(\det(\boldsymbol{\alpha}^{-1})=0\)
yield:
\begin{itemize}
  \item \emph{Low‐frequency limit}
        \(\omega,\omega_{c}\ll\omega_{p}\):\;
        a single resonance
        \(\omega_{0}=\omega_{c}\) with RHCP and LHCP dipoles that are
        degenerate in free space but split inside the cavity.
  \item \emph{High‐frequency limit}
        \(\omega\gg\omega_{c}\):\;
        the familiar triplet\cite{hadad2010magnetized}
        \begin{equation}
        \label{eq:gyro_triplet}
        \omega_{\pm}
          =\frac{\omega_{p}}{\sqrt3}
             \sqrt{1+\tfrac34\omega_{c}^{2}/\omega_{p}^{2}}
             \pm\frac{\omega_{c}}{2},
        \quad
        \omega'_{0}=\omega_{p}/\sqrt3 ,
        \end{equation}
        where \(\omega_{\pm}\) lie in the \(xy\)-plane and
        \(\omega'_{0}\) is \(z\)-polarised.
\end{itemize}

\subsubsection*{Results and Discussion}

\begin{figure*}[!t]
  \centering
  \includegraphics[width=0.85\textwidth]{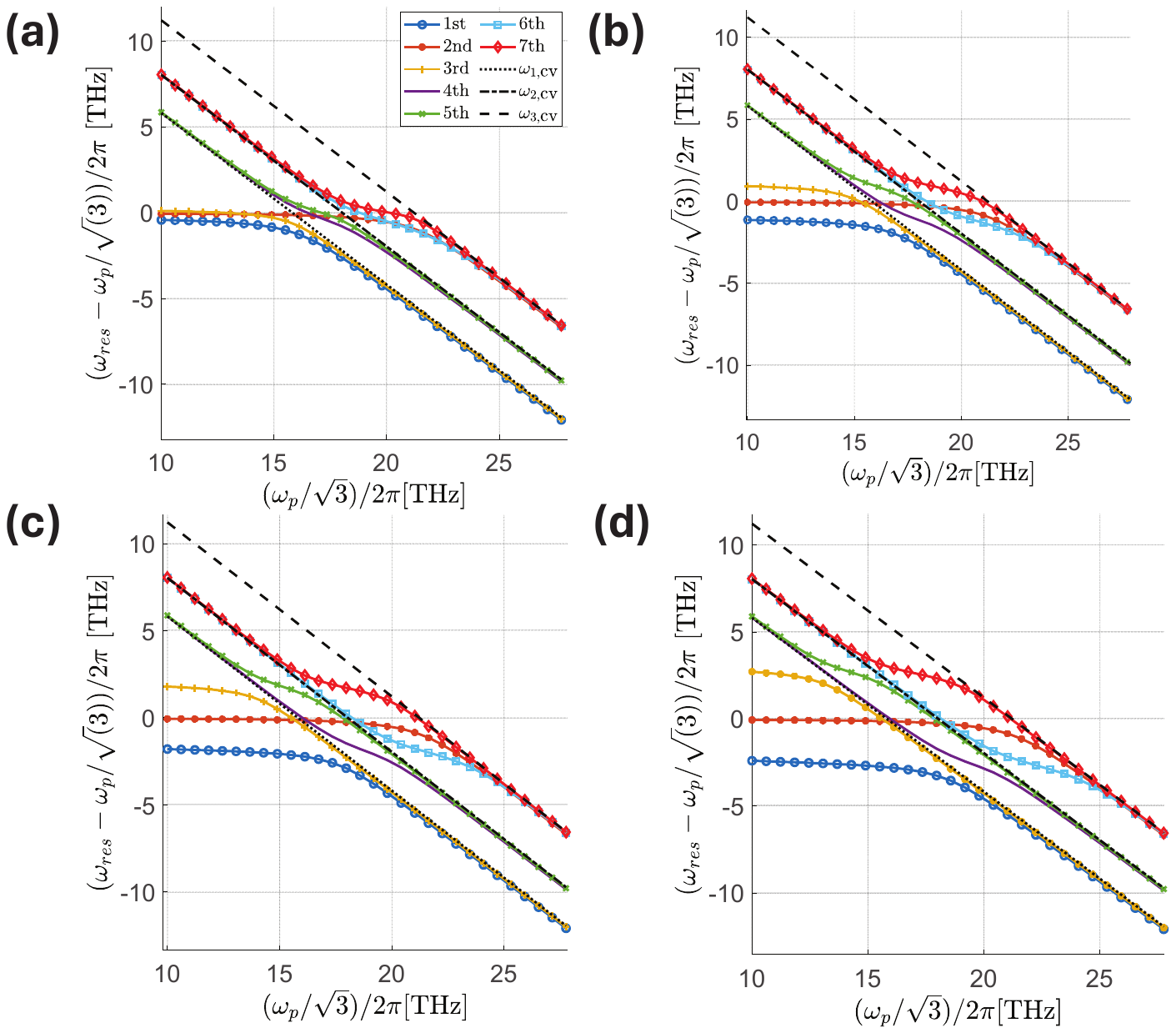}
  \caption{Coupled resonances versus plasma frequency
           \(\omega_{p}\) for four fixed cyclotron frequencies.
           Dashed lines mark the cavity cut-offs
           \(\omega_{-eps-converted-to.pdfi,\mathrm{cv}}\).}
  \label{fig:gyro_fb_all}
\end{figure*}

\paragraph*{Variation of \(\omega_{p}\) (fixed \(\omega_{c}\)).}
Figure~\ref{fig:gyro_fb_all} shows the six lowest coupled resonances
versus \(\omega_{p}\) for four fixed biases
\(\omega_{c}/2\pi=0.3,\,1.2,\,2.1,\,3~\text{THz}\).
With increasing bias the cavity‐dominated pairs (\#4–\#5 and
\#6–\#7) split, producing an “eye’’ around the unmagnetised
particle frequency \(\omega_{p}/\sqrt3\).
The axial dipole (\#2, brown curve) is almost insensitive to bias,
whereas the transverse pair (\#1 and \#3) separates in proportion
to~\(\omega_{c}\).

\begin{figure*}[!t]
  \centering
  \hspace{-2cm}
  \includegraphics[width=\textwidth]{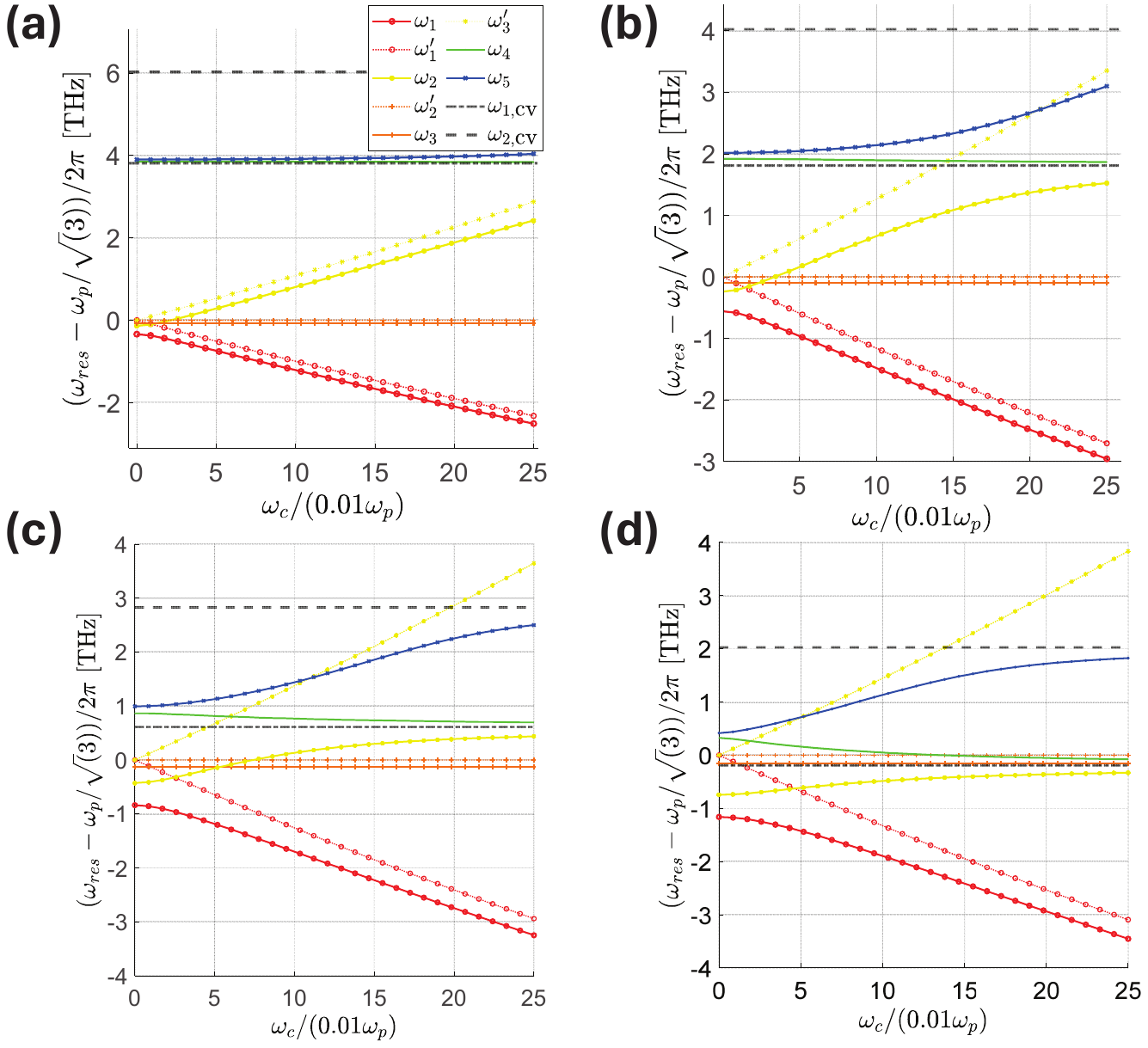}
  \caption{Coupled resonances versus cyclotron frequency
           \(\omega_{c}\) for four fixed plasma frequencies.
           Dotted lines: cavity cut-offs \(\omega_{i,\mathrm{cv}}\);
           dashed lines: isolated-particle roots
           \(\omega'_{i}\) from \eqsref{eq:gyro_triplet}.}
  \label{fig:gyro_fp_all}
\end{figure*}

\paragraph*{Variation of \(\omega_{c}\) (fixed \(\omega_{p}\)).}
Figure~\ref{fig:gyro_fp_all} fixes
\(\omega_{p}/2\pi=12,\,14,\,15.2,\,16~\text{THz}\) and sweeps the bias.
For \(\omega_{c}\ll\omega_{p}\) the isolated–sphere prediction
\begin{equation}
\label{eq:Taylor_gyro}
\omega_{\pm}\approx\frac{\omega_{p}}{\sqrt3}
               \pm\frac{\omega_{c}}{2}
               +\frac{\sqrt3}{8}\frac{\omega_{c}^{2}}{\omega_{p}},
\quad
\omega_{0}=\omega_{p}/\sqrt3
\end{equation}
remains accurate until the first cavity cut-off constrains mode~3.
The cavity thus acts as a spectral “rail’’: a mode born above (below)
a cut-off cannot cross it.

Overall, the static bias lifts the residual degeneracy of the
sphere–cavity system and generates a rich, non‐perturbative
reorganisation of its spectrum.

\subsection{Chiral (Bi-Isotropic) Sphere}
\label{Sec. Chiral}

\subsubsection*{Formulation}

Under the dipole approximation a point-like chiral sphere of volume
\(V\) is characterised by its dynamic polarizability tensor
\(\boldsymbol{\alpha}\), which maps the incident fields onto the
induced electric and magnetic moments\,\cite{Canaguier-Durand2015,mun2020electromagnetic}:
\begin{equation}
\begin{bmatrix}
  \mathbf p\\[2pt]
  \mathbf m
\end{bmatrix}
=\boldsymbol{\alpha}
\begin{bmatrix}
  \mathbf E_{\mathrm{inc}}(\mathbf r',\mathbf r')\\[2pt]
  \mathbf H_{\mathrm{inc}}(\mathbf r',\mathbf r')
\end{bmatrix}.
\end{equation}

\paragraph*{Energy-conserving polarizability.}
The quasi-static tensor \(\boldsymbol{\alpha}_{s}\) is augmented by the
radiation‐loss term,
\begin{subequations}
\begin{align}
\boldsymbol{\alpha}^{-1}&=
\boldsymbol{\alpha}_{s}^{-1}
+j\frac{k^{3}}{6\pi}
  \operatorname{diag}\!\bigl[\varepsilon_{0}^{-1}\mathbf I_3,
                            \mu_{0}^{-1}\mathbf I_3\bigr],\\
\boldsymbol{\alpha}_{s}&=
\begin{bmatrix}
  \boldsymbol{\alpha}_{ee} & -j\,\boldsymbol{\alpha}_{em}\\
  j\,\boldsymbol{\alpha}_{em}^{T} & \boldsymbol{\alpha}_{mm}
\end{bmatrix},
\end{align}
\end{subequations}
where \(\mathbf I_3\) is the \(3\times3\) identity.
The three constituent dyadics are
\begin{subequations}
\begin{align}
\boldsymbol{\alpha}_{ee}&=
   3\varepsilon_{0}V
   \frac{(\varepsilon_{r}-1)(\mu_{r}+2)-\kappa^{2}}
        {(\varepsilon_{r}+2)(\mu_{r}+2)-\kappa^{2}}
   \mathbf I_3,\\
\boldsymbol{\alpha}_{mm}&=
   3\mu_{0}V
   \frac{(\varepsilon_{r}+2)(\mu_{r}-1)-\kappa^{2}}
        {(\varepsilon_{r}+2)(\mu_{r}+2)-\kappa^{2}}
   \mathbf I_3,\\
\boldsymbol{\alpha}_{em}&=
   \frac{9\kappa V/\eta}
        {(\varepsilon_{r}+2)(\mu_{r}+2)-\kappa^{2}}
   \mathbf I_3,
\end{align}
\end{subequations}
with \(\eta\) the medium impedance, \(\kappa\) the chiral parameter,
and
\begin{subequations}
\begin{align}
\varepsilon_{r}(\omega)&=
   1-\frac{\omega_{p}^{2}}{\omega^{2}},\\
\mu_{r}(\omega)&=
   1+\frac{F\omega_{0}^{2}}{\omega_{0}^{2}-\omega^{2}},
\end{align}
\end{subequations}
where \(\omega_{p}\) is the plasma frequency,
\(\omega_{0}\) a Lorentz resonance, and \(F\) the so called `SRR filling factor', a measure for the chirality strength.

\paragraph*{Insertion into the cavity.}
Replacing \(\boldsymbol{\alpha}\) by the cavity-renormalized tensor using \eqsref{eq:sec to local}
\begin{equation}
\boldsymbol{\alpha}_{\mathrm{eff}}^{-1}=
  \boldsymbol{\alpha}^{-1} - \mathbf{G}_\mathrm{loc}(\mathbf{r}'), 
 \mathbf{G}_\mathrm{loc}(\mathbf{r}')=\underbrace{\begin{bmatrix}
     \mathbf G^{s}_{ee} & \mathbf G^{s}_{em}\\
     \mathbf G^{s}_{em} & \mathbf G^{s}_{mm}
   \end{bmatrix}}_{\mathbf G^{s}(\mathbf r',\mathbf r')}
\label{eq:alpha_eff_chiral}
\end{equation}
Then, the joint resonances of the chiral particle and the cavity are solved using \eqsref{eq:eig}.

\subsubsection*{Results}
\begin{figure}[!p]
  \centering
  \vspace{-1.5cm}
  \includegraphics[width=0.45\textwidth]{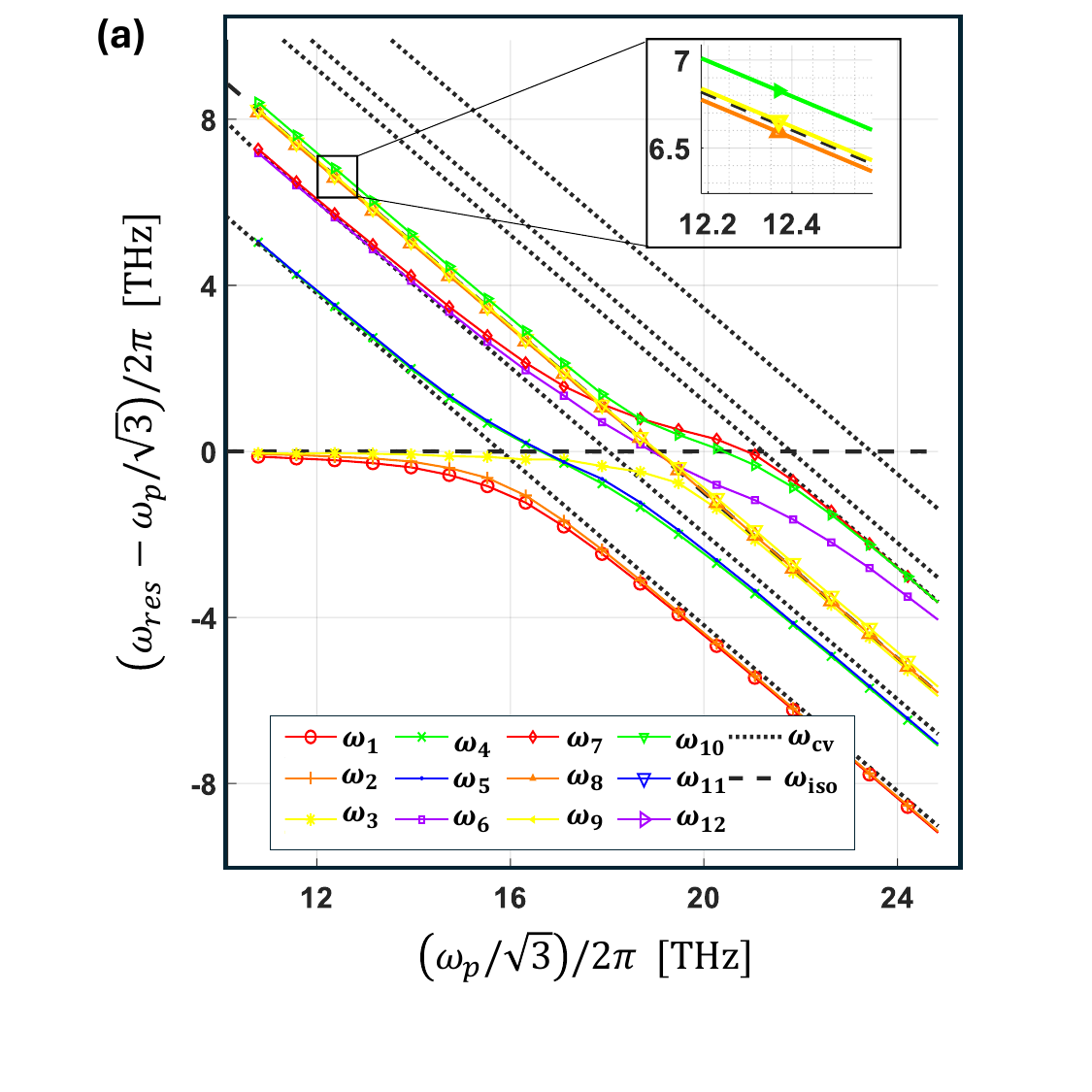}
  \\  \vspace{-0.5cm}
   \includegraphics[width=0.45\textwidth]{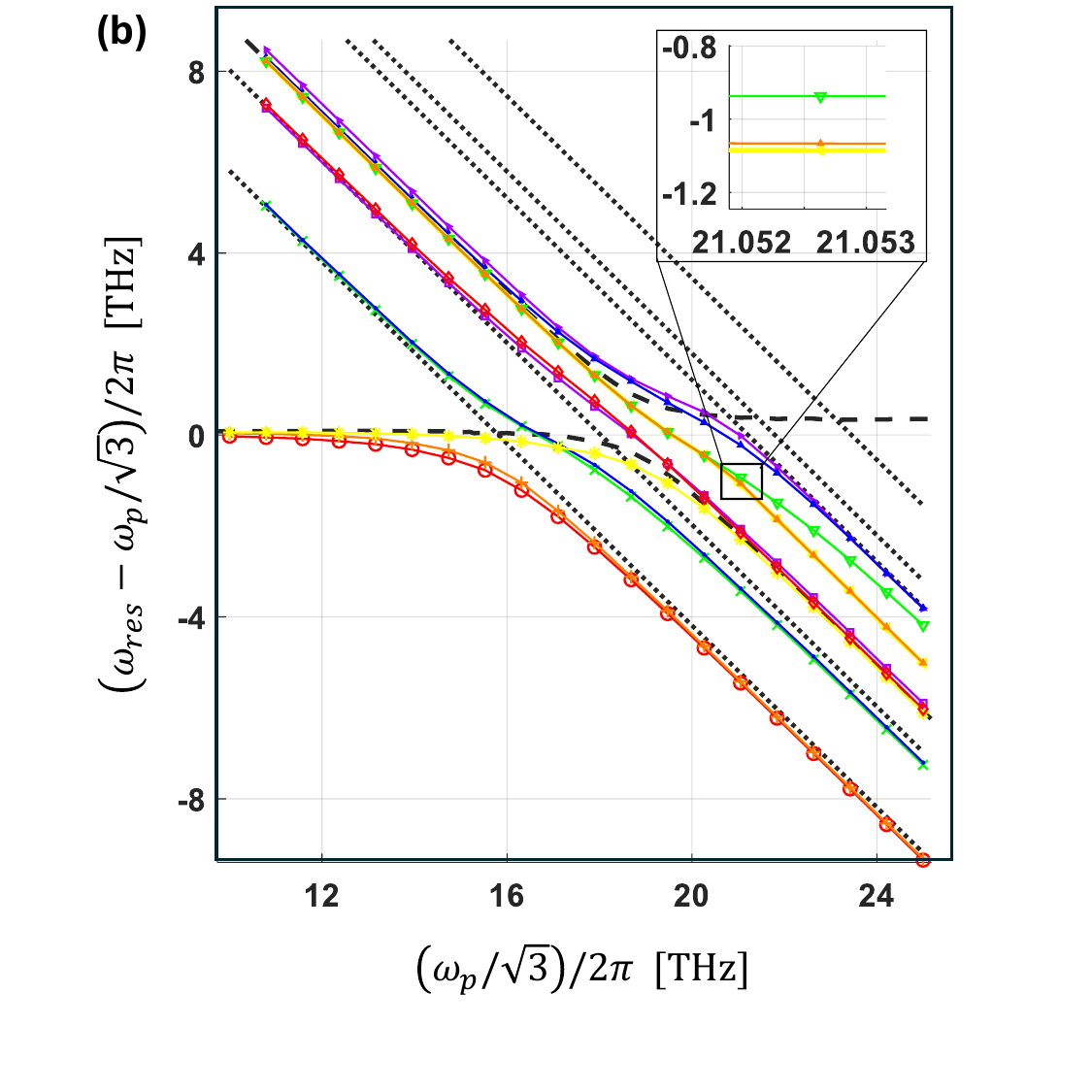}
     \\ \vspace{-0.5cm}
    \includegraphics[width=0.45\textwidth]{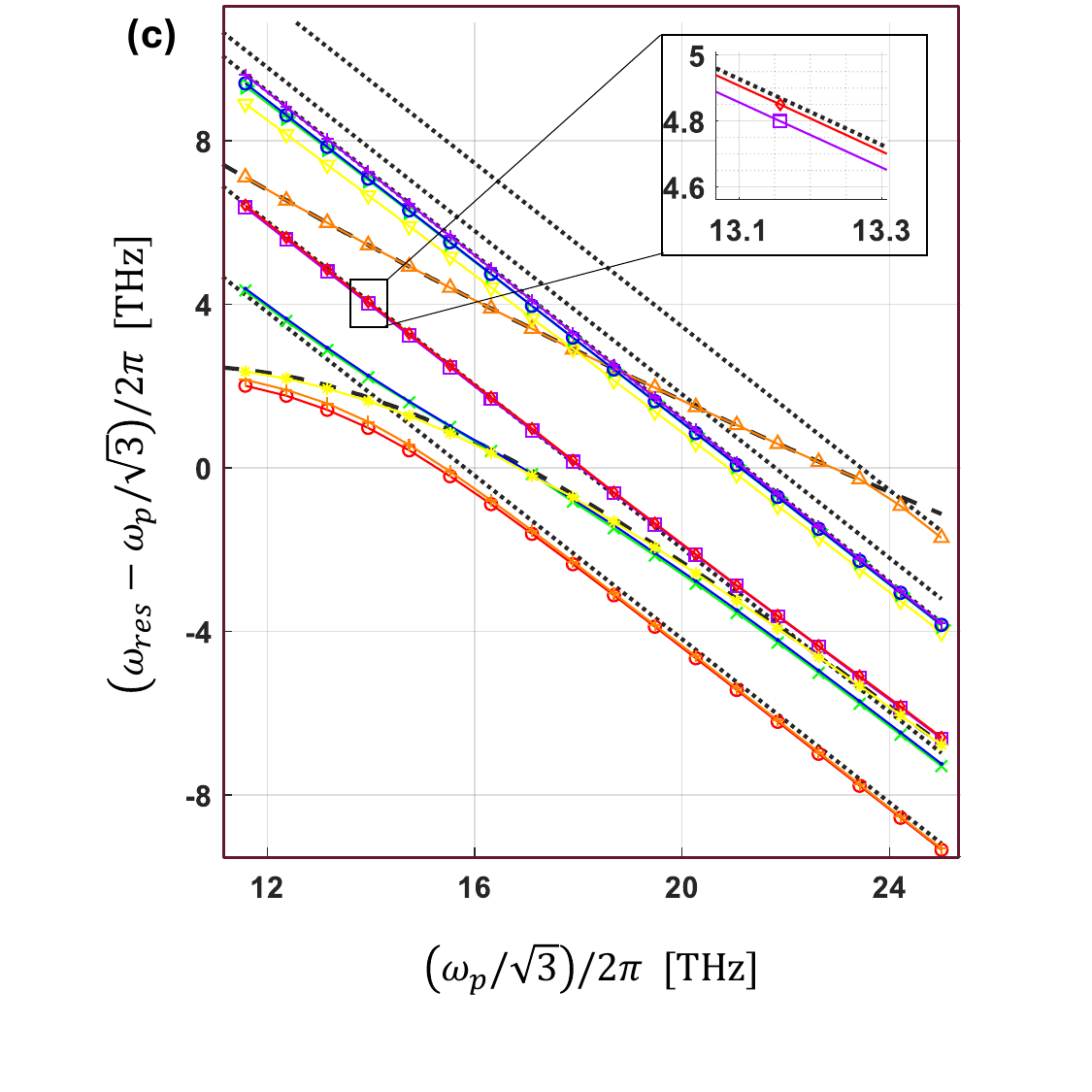}
      \vspace{-1cm}
  \caption{Coupled resonances versus plasma frequency
           \(\omega_{p}\) for three chiral parameters:
           (a)~\(\kappa=0\), (b)~\(\kappa=0.4\), (c)~\(\kappa=2\).
           Dotted lines: empty-cavity cut-offs
           \(\omega_{i,\mathrm{cv}}\);
           dashed lines: isolated-particle roots
           \(\omega_{e}\) (horizontal) and
           \(\omega_{m}\) (slanted).}
  \label{Chiral_response}
\end{figure}

Similarly to previous examples, we investigated the effect of the plasmonic frequencies on the coupled system. We analyze three cases with $\kappa=0$, $\kappa=0.4$, and $\kappa=2$ while fixing the resonance frequency parameter $\omega_0=17$THz, and F=0.6. Throughout the analysis, we distinguish again between the two families of resonances. \textit{cavity/particle} consists of two parts, {${\rm P_1}$},{${\rm P_2}$} where the former groups the first three resonances, and the latter groups of either the last three solutions for Fig.~\ref{Chiral_response}(a)) or the last five (Fig.~\ref{Chiral_response}(b-c)) solutions. More details are provided below. The second family \textit{particle/cavity} is also divided in two, {${\rm CV_1}$} consisting of $\omega_4,\omega_5$; {${\rm CV_2}$} consisting of $\omega_6,\omega_7$. (Fig.~\ref{Chiral_response}(a-c)).

For $\kappa=0$, the isolated particle has two resonance frequencies
\begin{align}
    &\omega_e=\frac{\omega_p}{\sqrt{3}} \\
    &\omega_m= \omega_0 \sqrt{\frac{3}{3-F}}
\end{align}

At the resonance $\omega_e$, only the electric dipole is excited, and at $\omega_m$, only the magnetic moment is excited.
Each has three degenerative states due to spherical symmetry. When embedded in the cavity, the cavity mediates the coupling between the electric dipole and the magnetic moment, and the resulting resonance frequencies are depicted in Fig.~\ref {Chiral_response}(a). In general, as demonstrated so far, the coupling of the particle to the cavity breaks the degeneracy in the system. For low plasmonic frequency $\omega_p$, {${\rm CV_1}$} and {${\rm CV_2}$} occur near the first $\omega_{1,\mbox{cv}} $ and second $\omega_{2,\mbox{cv}} $ (empty) cavity resonance, respectively and their dynamics as a function of increasing $\omega_p$ is similar to the isotropic case (Fig.~\ref{fig:epsart_summ}). Moreover, the resonance {${\rm P_1}$} and {${\rm P_2}$} occurs near $\omega_e$ (horizontal dash line) and $\omega_m$ (slanted dash line) respectively. On one hand, The migration of {${\rm P_1}$} is similar to when the particle was isotropic and modeled by a simple electric dipole (Fig.~\ref{fig:epsart_summ}), with the main difference that the 3rd resonance (yellow line) associated with the dipole in $\hat{z}$ axis migrates towards the $\omega_m$, as $\omega_p$ increases and for $\omega_p>\omega_0\sqrt{3}$). In the other, in the {${\rm P_2}$} group, only the resonance associated initially with the magnetic moment excited in $\hat{z}$ axis is strongly affected, and hereby migrates to the third cutoff $\omega_{3,\mbox{cv}} $. This behavior is analogous to one of the first two resonances in {${\rm P_1}$}.
Importantly, each of the remaining two resonances is degenerative with two states due to the inherited degeneracy in the cavity ($a=b$).

For $\kappa>0$, the isolated particle resonates predominantly as a magnetic moment. Although the particle exhibits two resonance frequencies, with $\omega_m$ one of its asymptotic values ( for $\kappa=0.6$, $\omega_e$ remains an asymptote) the curves of the frequencies with respect to $\omega_p$ do not cross, and the gap between the curves increases with $\kappa$. In the coupled system, the dynamics of {${\rm P_1}$}'s and {${\rm CV_1}$}'s frequencies follow a similar trend as in $\kappa=0$ case. Further, for both $\kappa=0.4$ and $\kappa=2$, the strong coupling of the excited magnetic moment to the cavity kept {${\rm CV_2}$}'s frequencies confined as they are led from the frequencies from $\omega_{2,\mbox{cv}} $ towards $\omega_m$. Concerning {${\rm P_2}$}'s frequency, we observe two interesting outcomes: first, the coupling breaks the cavity imposed degeneracy noted in the case of $\kappa=0$ therefore there is a total of five non-degenerative modes in {${\rm P_2}$}. Second, there is a gap between the {${\rm P_2}$} and {${\rm CV2}$}, similar to the one between the resonance of the isolated particle. For $\kappa=0.4$, the {${\rm P_2}$}'s frequencies migrate from the $\omega_m$ towards the 3rd cavity resonance frequency while for  $\kappa=2$, they are mostly confined near the third cavity resonance frequency due to the increased gap.

Overall, chirality removes the last degeneracies, enlarges the
electric–magnetic spectral gap, and locks selected modes to the upper
cut-off when the optical activity is strong, all these fine details are captured by the suggested exact regularization approach.


\section{Conclusion}\label{Sec:conc}

This work examined the mutual interaction between a resonant particle
and a rectangular cavity by combining the discrete–dipole
approximation with polarizability theory.
Two main contributions were established.

\begin{enumerate}
\item \textbf{Semi-analytical local-field evaluation.}  
      We introduced an exact yet computationally light scheme for
      calculating the \emph{local} field that drives the particle
      dipole.
      The scheme relies on a ladder-type subtraction of
      Green-function singularities, so that a standard free-space
      radiation correction suffices for any tabulated
      polarizability.  
      The same subtraction hierarchy remains applicable when the
      cavity modes are known only numerically, a regime where Ewald
      summation or image methods are inapplicable.

\item \textbf{Physical insight into collective resonances.}  
      The coupled spectrum splits into two families:
      (i) \emph{particle-like} modes shifted by the cavity loading and
      (ii) \emph{cavity-like} modes perturbed by the particle.  
      Mutual loading lifts the degeneracy inherent in the isolated
      subsystems.  
      The framework accommodates arbitrary material tensors; we
      demonstrated isotropic, gyrotropic, and chiral spheres.  
      In an accompanying paper\,\cite{Sadzi2024b} the gyrotropic case
      is explored further, revealing a magnetisation–loss threshold
      that governs Lorentz (non-)reciprocity.
\end{enumerate}

\section*{Acknowledgment}

This work was supported in part by the Israel Science Foundation under
Grant 1457/23.  
The authors thank Prof.~Vitaliy Lomakin and Prof.~Sergei Tretyakov for
insightful discussions and constructive comments.


\appendix
\counterwithin{equation}{section}

\section*{Local Dyadic Green’s Function for a \texorpdfstring{$\hat{\mathbf z}$}{z}-Oriented Magnetic Dipole}
\label{sec:level_3b}

This appendix mirrors the three–step subtraction procedure of
Sec.~\ref{sec:localGF}, but for a \emph{magnetic} point-current
source
\(\mathbf M=M_{0}\,\delta(\mathbf r-\mathbf r')\hat{\mathbf z}\)
(\(M_{0}=1\,[\mathrm{V\,m}]\)).
The algebra closely parallels the electric case; only the modal
sources and the resulting transmission-line (TL) data change.
Unless stated otherwise, the TL models for E- and H-modes differ only
through the characteristic impedance \(Z_{i}\).
Throughout, $\varsigma_{z,z'}=\operatorname{sgn}(z-z')$ and
$k_{t,i}$, $\kappa_{i}$, $Z_{i}$ follow the same definitions as in
Sec.~\ref{sec:localGF}, where $i$ denotes the spectral two-index, $mn$, $m\xi$, and $\xi\eta$, for the three subtraction steps in \eqsref{eq:ladder}.

\subsection{Step I: Cavity Minus Infinite Rectangular Waveguide}
\label{subsec:mag_StepI}

Figures \ref{fig:cavity}(a)–(d) still describe the geometry, the only
difference being the magnetic source.
Because a magnetic dipole is dual to an electric dipole, only
\emph{H-modes} are now excited, and the modal voltage source in Fig.~\ref{fig:cavity} is replaced with a dual shunt current source (the same polarization as in the TL models in Fig.~\ref{fig:rect})

\paragraph*{Modal sources.}
Using the same transverse eigenfunctions
\(\Psi_{mn}(x,y)\) in~\eqsref{eq:Psi},
the H-modes modal voltages and currents are \cite{felsen2001radiation}
\begin{subequations}
\begin{align}
&v_{mn}(z;\mathbf r') = 0,\\ 
&i_{mn}(z;\mathbf r') =
   j\frac{k_{t,mn}}{\omega\mu_{0}}
   \Psi_{mn}(x',y')\,\delta(z-z').
\end{align}
\end{subequations}
(the E-modes sources can be readily shown to be identically zero, as expected physically)
\paragraph*{TL Green functions.}
Solving the short-circuited TL (cavity) and the open TL (RG1) gives

\begin{subequations}
\begin{align}
&\mathcal V_{\mathrm{CV},mn}(z,z')%
  =\frac{Z''_{mn}}{2\!\left(e^{-2j\kappa_{mn}c}-1\right)}
   \bigl[e^{-j\kappa_{mn}|z-z'|}\notag\\[-2pt]
& -e^{-j\kappa_{mn}(z+z')}
      -e^{-j\kappa_{mn}\!\left(2c-(z+z')\right)}
      +e^{-j\kappa_{mn}\!\left(2c-|z-z'|\right)}\bigr],\\[4pt]
&\mathcal I_{\mathrm{CV},mn}(z,z')%
  =\frac{1}{2\!\left(e^{-2j\kappa_{mn}c}-1\right)}
   \bigl[\varsigma_{z,z'}e^{-j\kappa_{mn}|z-z'|}-\notag\\
& e^{-j\kappa_{mn}(z+z')}
     \! +\!e^{-j\kappa_{mn}\!\left(2c-(z+z')\right)}
      \!-\!\varsigma_{z,z'}e^{-j\kappa_{mn}\!\left(2c-|z-z'|\right)}\bigr],\\[4pt]
&\mathcal V_{\mathrm{RG1},mn}(z,z') = -\tfrac12 Z''_{mn}\,e^{-j\kappa_{mn}|z-z'|},\\
&\mathcal I_{\mathrm{RG1},mn}(z,z') = -\tfrac12 \varsigma_{z,z'}\,e^{-j\kappa_{mn}|z-z'|}.
\end{align}
\end{subequations}
Note that these 1D Green's functions correspond to essentially identical TL models as in Fig.~\ref{fig:rect}, and thus are identical (up to the use of the proper characteristic line impedance) to those given in \eqsref{eq:VIRG2} and \eqsref{eq:VIPP1}. 

\paragraph*{Subtracted fields.}
With
\(\Delta\bar{\mathcal V}^{(1)}_{mn}=2Y''_{mn}
 (\mathcal V_{\mathrm{CV},mn}-\mathcal V_{\mathrm{RG1},mn})\)
and
\(\Delta\bar{\mathcal I}^{(1)}_{mn}=2
 (\mathcal I_{\mathrm{CV},mn}-\mathcal I_{\mathrm{RG1},mn})\),
the Step I dyadic entries are

\begin{subequations}
 \begin{align}
    E_{xz}^{(1)}(\bm{r,r'})=& \,\frac{j}{ab}
    \sum_i \frac{\epsilon_n \epsilon_m  }{2}\Delta\bar{\mathcal{V}}_{mn}^{(1)}\,
    \,
   \frac{k_y}{\kappa}\,
   \times\notag \\ & C_x(x') C_y(x') C_x(x) S_y(y)
\\
    E_{yz}^{(1)}(\bm{r,r'})=& \,\frac{-j}{ab} \sum_i \frac{\epsilon_n \epsilon_m  }{2}  \Delta\bar{\mathcal{V}}_{mn}^{(1)}\,
    \frac{k_x}{\kappa}\,\times\notag \\ &
    C_x(x') \, C_y(y') S_x( x) \,C_y(y)
\\
    E_{zz}^{(1)}(\bm{r,r'})= & 0
\\
    H_{xz}^{(1)}(\bm{r,r'})=& \, \frac{j}{\omega \mu_0\,ab}
    \sum_i \frac{\epsilon_n \epsilon_m  }{2}
    \Delta\bar{\mathcal{I}}_{mn}^{(1)}\
    \,
    k_x\,
    \;
  \times\notag \\ & C_x(x') C_y(y') S_x(x) \, C_y(y)
\\
    H_{yz}^{(1)}(\bm{r,r'})=& \, \frac{j}{\omega \mu_0 \, ab}
    \sum_i \frac{\epsilon_n \epsilon_m  }{2}
   \Delta\bar{\mathcal{I}}_{mn}^{(1)}
    \,
    k_y \,
   \times\notag \\ & C_x(x') C_y(y') C_x(x)\,S_y(y)
\\
    H_{zz}^{(1)}(\bm{r,r'})=& \,\frac{1}{\omega \mu_0 \, ab} \sum_i \frac{\epsilon_n \epsilon_m  }{2}
   \Delta\bar{\mathcal{V}}_{mn}^{(1)}\,
   \frac{k_{t,mn}^2 }{\kappa}
   \times\notag \\ & C_x(x') C_y(y') C_x(x) C_y(y)
\end{align}
\end{subequations}

\subsection{Step II: Alternative Waveguide Representation Minus PP Waveguide}
\label{subsec:mag_StepII}

Figures \ref{fig:rect}(a)–(d) apply unchanged except for the modal sources that are now should be the dual of the electric source problem. Namely, the shunt current source will be replaced with a series voltage source in the same polarization as in the TL models in Fig.~\ref{fig:cavity}. 

\paragraph*{Modal sources.}
Using the eigenfunctions \(\Phi_{m,\xi},\Psi_{m,\xi}\)
of~\eqsref{eq:PhiPsi_RG2}, the only non-zero TL source is a \emph{voltage}
\begin{equation}
v_{m\xi}(y;\mathbf r')=
  \begin{cases}
     \displaystyle
     \frac{-1}{k_{t,m\xi}}
     \partial_x\Phi^*_{m\xi}(x',z')\delta(y-y'),
     &\text{E-mode},\\[6pt]
     \displaystyle-
     \frac{1}{k_{t,m\xi}}
     \partial_z\Psi^*_{m\xi}\delta(y-y'),
     &\text{H-mode}.
  \end{cases}
\end{equation}

\paragraph*{TL Green functions.}
The short-circuited TL (RG2) and the infinite TL (PP1) yield
\(\mathcal V_{\mathrm{RG2},m\xi},\mathcal I_{\mathrm{RG2},m\xi}\)
in~\eqsref{eq:VCV} and
\(\mathcal V_{\mathrm{PP1},m\xi},\mathcal I_{\mathrm{PP1},m\xi}\)
in~\eqsref{eq:TL_RG1}, with the proper change of voltages and currents names, coordinate $z\leftrightarrow y$, $b\leftrightarrow c$ and the characteristic impedance.

\paragraph*{Subtracted fields.}
Define
\(\Delta\bar{\mathcal V}^{(2)}_{m\xi}=2(\mathcal
V_{\mathrm{RG2},m\xi}-\mathcal V_{\mathrm{PP1},m\xi})\) and
\(\Delta\bar{\mathcal I}^{(2)}_{m\xi}=2Z_{m\xi}(\mathcal
I_{\mathrm{RG2},m\xi}-\mathcal I_{\mathrm{PP1},m\xi})\).
Using the same trigonometric shorthand as before, the Step II entries
become
\begin{subequations}
\label{eq:mag_StepII}
\begin{align}
E_{xz}^{(2)}
 &=-\frac{1}{2\pi a}
   \sum_{m\neq0}\frac{\epsilon_m}{2}\!C_{x}(x)C_{x}(x')\times\notag\\&\quad
   \int_{\xi}\!\!
   \,\Delta\bar{\mathcal V}^{(2)}_{m\xi}e^{-j\xi(z-z')}d\xi,\\
E_{yz}^{(2)}
 &=-\frac{j}{2\pi a}
   \sum_{m\neq0}\!C_{x}(x')S_{x}(x)\times\notag\\&\quad
   \int_{\xi}\!\!
   \frac{k_x}{\kappa_{m\xi}}\Delta\bar{\mathcal I}^{(2)}_{m\xi}e^{-j\xi(z-z')}d\xi,\\
E_{zz}^{(2)}&=0, \\   
H_{xz}^{(2)}
 &= \frac{j}{2\pi a \omega\mu_0}
    \sum_{m\neq0}\!C_{x}(x')S_{x}(x)\times\notag\\&\quad
    \int_{\xi}\!\!
    \frac{\xi k_{x}}{\kappa_{m\xi}}\Delta\bar{\mathcal I}^{(2)}_{i}e^{-j\xi(z-z')}d\xi,\\
H_{yz}^{(2)}
 &= \frac{-1}{2\pi a \omega \mu_0}
    \sum_{m\neq0}\!\frac{\epsilon_m}{2}C_{x}(x')C_{x}(x)\times\notag\\&\quad
    \int_{\xi}\!\!
    \Delta\bar{\mathcal V}^{(2)}_{m\xi}e^{-j\xi(z-z')}\xi d\xi,\\
H_{zz}^{(2)} & = \frac{1}{2\pi a \omega\mu_0} \sum_{m\neq0} \frac{\epsilon_m}{2}C_x(x')C_x(x)\times\notag\\&\quad\int_{\xi} \frac{\kappa_{m\xi}^2+k_x^2}{\kappa_{m\xi}}\Delta\bar{\mathcal{I}}^{(2)}_{m\xi} e^{-j\xi(z-z')}.    
\end{align}
\end{subequations}

\subsection{Step III: Alternative PP Representation Minus Free Space}
\label{subsec:mag_StepIII}

Figures \ref{fig:pp}(a)–(d) again apply after the duality swap of the shunt current source with a series voltage source. Consequently, the 1D Green functions are as in Step II, besides trivial replacements $x\!\leftrightarrow\!y,\;a\!\leftrightarrow\!b$. Yielding $\bar{\mathcal{V}}_{PP2}^{(3)}(x,x'), \bar{\mathcal{V}}_{FS}^{(3)}(x,x'),\bar{\mathcal{I}}_{PP2}^{(3)}(x,x'), \bar{\mathcal{I}}_{FS}^{(3)}(x,x')$.
The modal eigenfunctions correspond to the cross section rather than to the source, and are given by \eqsref{eq:modal fs}. 
Consequently, the modal sources for E- and H- modes read, 
\begin{equation}
v_{\xi\eta}(y;\mathbf r')=
  \begin{cases}
     \displaystyle
     \frac{-1}{k_{t,\xi\eta}}
     \partial_y\Phi^*_{\xi\eta}(y',z')\delta(x-x'),
     &\text{E-mode},\\[6pt]
     \displaystyle
     \frac{-1}{k_{t,\xi\eta}}
     \partial_z\Psi^*_{\xi\eta}(y',z')\delta(x-x'),
     &\text{H-mode}.
  \end{cases}
\end{equation}

Define \(\Delta\bar{\mathcal V}^{(3)}_{\xi\eta}=2(\mathcal V_{\mathrm{PP2},\xi\eta}-\mathcal V_{\mathrm{FS},\xi\eta})\),
\(\Delta\bar{\mathcal I}^{(3)}_{\xi\eta}=2Z_{\xi\eta}(\mathcal I_{\mathrm{PP2},\xi\eta}-\mathcal I_{\mathrm{FS},\xi\eta})\).
With $(\bar\rho,\bar\phi)$ the polar coordinates of
$(z-z',y-y')$, the Step III contributions are
\begin{subequations}
\begin{align}
    E_{xz}^{(3)} (\bm{r,r'})=&\frac{j}{4\pi}
    \, \sin(\bar{\phi})
     \int_{\infty}
    dq \; \frac{q^2}{\kappa_{\xi\eta}}
    \,
    \mathcal{J}_1(q\bar{\rho})
    \,
    \Delta\bar{\mathcal{I}}_{\xi\eta}^{(3)}(x,x')
\\
    E_{yz}^{(3)}  (\bm{r,r'})=&  \frac{1}{4\pi}
     \int_{\infty}
    dq \; q
    \, \mathcal{J}_0(q\bar{\rho})   \,
    \Delta\bar{\mathcal{V}}_{\xi\eta}^{(3)}(x,x')
\\
    E_{zz}^{(3)} (\bm{r,r'})=&0
\\    H_{xz}^{(3)} (\bm{r,r'})=&\frac{j}{4\pi}
    \frac{\cos(\bar{\phi})}{\omega \mu_0}\;
     \int_{\infty}
    dq \;
    q^2
    \mathcal{J}_1(q\bar{\rho})\;
    \Delta\bar{\mathcal{V}}_{\xi\eta}^{(3)}(x,x')
\\
  H_{yz}^{(3)} (\bm{r,r'})=&\frac{1}{8\pi}
  \frac{\sin(2\bar{\phi})}{\omega \mu_0}
 \int_{\infty}
     dq\,\frac{q^3}{\kappa_{\xi\eta}}
    \mathcal{J}_2(q \bar{\rho})
    \,
    \Delta\bar{\mathcal{I}}_{\xi\eta}^{(3)}(x,x')
\\
    H_{zz}^{(3)} (\bm{r,r'})=&  \frac{1}{8\pi \, \omega\mu_0}
    \int_{\infty}
    dq \; \bigl[
    \frac{q^3}{\kappa_{\xi\eta}} \mathcal{J}_2(q\bar{\rho})\,
    \cos(2\bar{\phi})\,   \notag\\&+
    \frac{q}{\kappa_{\xi\eta}}  \left(2k_0^2-q^2\right)  \mathcal{J}_0(q\bar{\rho})\bigr]
  \Delta\bar{\mathcal{I}}_{\xi\eta}^{(3)}(x,x')
\end{align}
\end{subequations}

\subsection{Assembly of the Local Green Dyadic}

Adding the three regularised contributions gives, for example,
\begin{subequations}
\begin{align}
\mathbf G^{s,he}_{xz}(\mathbf r,\mathbf r')
 &= \hat{\mathbf x}\hat{\mathbf z}\,
    \bigl(E_{xz}^{(1)}+E_{xz}^{(2)}+E_{xz}^{(3)}\bigr),\\
\mathbf G^{s,hh}_{xz}(\mathbf r,\mathbf r')
 &= \hat{\mathbf x}\hat{\mathbf z}\,
    \bigl(H_{xz}^{(1)}+H_{xz}^{(2)}+H_{xz}^{(3)}\bigr).
\end{align}
\end{subequations}
Setting $\mathbf r\!=\!\mathbf r'$ converts the secondary dyadic into
the \emph{local} Green function required for the cavity–particle
interaction analysis.


\bibliographystyle{IEEEtran}

\end{document}